\documentclass[12pt]{article}

\usepackage[english]{babel}
\usepackage{rotating}
\usepackage{authblk}
\usepackage{url}
\usepackage[letterpaper,top=2.5cm,bottom=2.5cm,left=2.5cm,right=2.5cm,marginparwidth=1.75cm]{geometry}

\usepackage{amsmath}
\usepackage{ amssymb }
\usepackage{graphicx}
\usepackage{xcolor, colortbl}

\usepackage{subfigure}
\usepackage{lscape}
\usepackage{array, multirow, multicol}
\usepackage[colorlinks=true, allcolors=blue]{hyperref}
\usepackage{fancyhdr}
\usepackage{cleveref}
\usepackage[backend=biber, sorting=none]{biblatex} 
\usepackage{mathtools}
\usepackage{cancel}
\usepackage{dsfont}
\usepackage{orcidlink}
\newcommand{\eq}[1]{\begin{align}#1\end{align}}

\renewcommand{\u}[1]{\textrm{U}(#1)}
\newcommand{\su}[1]{\textrm{SU}(#1)}

\newcommand{\nn}{\nonumber} 

\newcommand{\ba}{\begin{array}}
\newcommand{\ea}{\end{array}}

\newcommand{\lag}{{\cal L}}

\newcommand{\hc}{{\rm h.c.}}
\newcommand{\elev}[1]{^{#1}}
\hypersetup{colorlinks, citecolor=blue, linkcolor=newred, urlcolor=blue}
\definecolor{newred}{rgb}{0.68, 0.15, 0.20}
\definecolor{newgray}{rgb}{.85,.85, .85}

\DeclareUnicodeCharacter{2212}{-}

\title{Axion dark matter from extended misalignment with a constant-$\omega_\phi$ pre-oscillatory phase and dark radiation}

\author[]{José María Pérez-Poyatos} 
\affil[]{\textit{Universidad de C\'ordoba, E-14071 C\'ordoba, Spain}}
\affil{\textit{E-mail: }jm.perez@uco.es}
\date{}
\begin{document}
\maketitle

\begin{abstract}

 In this work, we extend the standard pre-inflationary misalignment mechanism for axion-like particles (ALPs) by introducing a pre-oscillatory phase with constant equation of state $\omega_\phi\in[-1,1]$, supported by the form of the scalar potential and, in general, suitable initial conditions. During the radiation-dominated era, the potential undergoes a rapid transition to the conventional cosine potential. The resulting change in the potential energy across the transition can drive the ALP into a kinetic misalignment phase ($\omega_\phi=1$) prior to the onset of oscillations. Motivated by persistent cosmological tensions, such as those in $H_0$ and $S_8$, we also investigate an ALP coupling to a dark radiation sector (DR), allowing for its decay. Using a Bayesian analysis, we constrain the ALP parameter space with current cosmological data. Our analysis shows that ALP-induced DR does not resolve the existing tensions. Instead, the data place robust constraints on the model, favoring negative values of $\omega_\phi$, establishing an upper bound on the pre-transition energy scale relative to the maximum of the cosine potential energy, and constraining the symmetry-breaking scale to $f_\phi\in[80,1.5\times10^{10}]~\mathrm{TeV}$, corresponding to ALP masses in the range $m_\phi\in[10^{-20},10^{-2}]~\mathrm{eV}$.
\end{abstract}

\section{Introduction}

Axions and axion-like particles (ALPs) are compelling candidates for dark matter. In addition, their cosmological production may be linked to phase transitions in the early Universe that modify the vacuum structure and can generate observable gravitational wave signals.

The QCD axion was initially proposed to resolve the Strong CP problem, addressing the absence of CP violation in strong interactions. To tackle this issue, an extension to the Standard Model (SM) gauge group, $\su3_c\times\su2_L\times \u1_Y$, is introduced by incorporating a global $\u1$ group \cite{Peccei:1977hh,Wilczek:1977pj,Weinberg:1977ma,Dine:1981rt}. This extension undergoes spontaneous breaking facilitated by the vacuum expectation value of a complex scalar field, singlet under the SM gauge symmetries,  at the energy scale $f_\phi$. Consequently, the axion arises as the (CP-odd) Goldstone boson linked to this breaking mechanism.

However, the global \u1 is broken at the quantum level, and the axion develops small (and temperature-dependent) mass and anomalous couplings to the SM gauge fields. Regarding fermions, the axion's couplings are derivative, thereby preserving the Goldstone nature of the axion. These couplings emerge at dimension 5 and are suppressed by the scale $f_\phi$.

The scale of spontaneous symmetry breaking $f_\phi$ was initially proposed to be around the electroweak scale $v\sim 246$~GeV. However, the lack of signals of the axion implies that it must be very weakly coupled to the SM particles, which in turn sets $f_\phi>10^9$~GeV: the so called invisible axion paradigm (see current bounds and future detection prospects~\cite{cajohare,Eby:2024mhd}). 

On the other hand, models that try to address other theoretical problems including Composite or Little Higgs~\cite{DelAguila:2019xec,Illana:2021uwu,Illana:2022wqv}, Extra Dimensions~\cite{Liu:2017gcn}, Majorons ~\cite{Berezhiani:1989fp,Cuesta:2021kca,Cuesta:2023awo}, or, more generally, String Theory~\cite{Svrcek:2006yi}, also include in their spectrum scalar particles resulting from the spontaneous breaking of global symmetry groups that exhibit properties similar to those of the QCD axion: they are singlets under the SM gauge group, naturally lightweight, and weakly coupled. They are commonly referred to as \textit{axion-like particles} (ALPs).

The intrinsic properties of ALPs render them promising candidates for addressing cosmological and astrophysical issues such as inflation (in the natural inflation scenario), protect the inflationary potential from dangerous quantum corrections \cite{Freese:1990rb,Adams:1992bn,Kim:2004rp,Dimopoulos:2005ac,Maleknejad:2011jw,Adshead:2012kp,Peloso:2015dsa}, explain particle production in the early Universe
\cite{Anber:2009ua,Cook:2011hg,Barnaby:2011qe,Namba:2015gja,Dimastrogiovanni:2016fuu,Peloso:2016gqs,Garcia-Bellido:2016dkw,Domcke:2016bkh}, the origin of pulsar timing array measurements \cite{Unal:2023srk,Guo:2023hyp} and the dark matter (DM) problem, see e.g.~\cite{Marsh:2015xka}. For a solution to the DM problem not involving axions but a mirror SM world see \cite{Oikonomou:2024geq}.

Various mechanisms exist for producing DM from ALPs. In this study, we confine our focus to non-thermal dark matter production mechanisms within the pre-inflationary framework, where the global symmetry responsible for ALP emergence is broken before inflation and remains so. Non-thermal mechanisms present an appealing alternative to the conventional thermal Weakly Interacting Massive Particle (WIMP) paradigm. A notable non-thermal mechanism is the so called realignment or conventional misalignment mechanism~\cite{Preskill:1982cy, Dine:1982ah,Abbott:1982af,Kolb:1990vq}. In the early Universe, ALPs may deviate from the CP-even minimum of their potential. When the Hubble parameter $H$ becomes comparable to the ALP mass $m_\phi$ at the temperature $T_*$ determined by $m_\phi=3H(T_*)$, the ALP initiates oscillations around the minimum. These coherent oscillations behave as pressureless matter and may reproduce the observed DM abundance, even in non-standard cosmologies \cite{Visinelli:2009kt,Visinelli:2017imh}. Pre-inflationary mechanisms circumvent issues like cosmic strings and domain walls, which are characteristic of the post-inflationary scenario \cite{Sakharov:1994id,Khlopov:1998uj,Khlopov:1999tm}. Nevertheless, they are subject to isocurvature constraints~\cite{Axenides:1983hj,Steinhardt:1983ia,Steinhardt:1984jj,Linde:1985yf,Seckel:1985tj,LYTH1990408,Turner:1990uz,Sikivie:2006ir,Beltran:2006sq,Kawasaki:2008sn,Kawasaki:2014una,Chen:2023txq}. 

Another mechanism that recently got more attention is the so called \textit{kinetic misalignment} (KM) mechanism \cite{Co:2019jts}. In this scenario the ALP has a large initial velocity in field space in the early Universe. The expansion of the Universe acts as a friction term, redshifting the ALP velocity. If the ALP kinetic energy still dominates over the potential energy during radiation domination era at the temperature $T_*$, the ALP can still explore several potential minima until kinetic energy redshifts to the height of the potential barrier, at which point it starts to oscillate. As in the conventional misalignment mechanism, the oscillations are responsible of generating DM.

On the other hand, Cosmology is currently living a high precision era. As a consequence, some discrepancies are emerging between different datasets. Probably one of the best known is the $H_0$ tension, an $\sim 5\sigma$ discrepancy between inferences of the today Hubble expansion rate using early and late time Universe datasets. This is largely driven by the discrepancy between the local measurement of the SH0ES collaboration using supernovas \cite{Verde:2019ivm,Knox:2019rjx,DiValentino:2021izs,Shah:2021onj} and
the model-dependent inference from the CMB and/or large-scale structure data using the $\Lambda\rm CDM$ model. A less stringent discrepancy is the $S_8$ tension, a discrepancy of $\sim 2-3 \sigma$ in the amount of clustering of matter
seen between late-time Universe datasets and the CMB
data \cite{Troster:2019ean,Ivanov:2019pdj,KiDS:2020suj,Hildebrandt:2020rno,KiDS:2020ghu,DES:2021wwk,Philcox:2021kcw,Krolewski:2021yqy,Garcia-Garcia:2024gzy,Nunes:2021ipq}. 

The aim of this work is twofold. First, assuming a constant ALP equation of state prior to a rapid transition in the ALP potential to the usual cosine-type form and allowing for a coupling to a dark radiation (DR) sector, we constrain the ALP parameter space using recent cosmological data. Second, we investigate whether this framework can address the persistent $H_0$ and $S_8$ tensions.

The paper is structured as follows: in Section~\ref{review} we review the different mechanisms for misalignment proposed in the literature, as well as the cosmological tensions. In Section~\ref{DRproduction} we present our model in detail, at both the background and perturbation levels, providing analytical expressions for the production of dark radiation generated by the oscillations of the ALP around one of the minima of its potential. In Section~\ref{Bayesiananalysis} we perform a Bayesian analysis to confront the ALP decaying model to current cosmological data to constrain the ALP parameter space and explore if it can indeed solve the $H_0$ and $S_8$ tensions. Finally, Section~\ref{conclusions} summarizes our findings and presents our conclusions.

\section{Review on misalignment mechanisms  and cosmological tensions}\label{review}

In this section, we review the various misalignment mechanisms proposed for axion-like particles (ALPs) as a candidate for dark matter. These mechanisms are crucial for explaining how ALPs can acquire the correct relic density in the pre-inflationary scenario. We also discuss the cosmological tensions that arise, particularly in relation to the Hubble constant and the matter power spectrum. ALPs may offer potential resolutions or deepen these tensions depending on their interactions with other fields and their production history.

\subsection{Misalignment mechanisms}

In this section we briefly review some of the different misalignment mechanisms that have been proposed in the literature to produce ALP DM in the pre-inflationary scenario. These mechanisms are highly non-thermal. After the spontaneous breaking of the global symmetry that gives rise to the ALP, this chooses (probably by some stochastic process) a initial value $\phi_i$ in each Hubble patch. Later on, inflation homogenizes the value of the ALP in the whole Universe. After inflation, the dynamics of the ALP is model dependent. However, when the Hubble parameter becomes comparable to the ALP mass, the ALP oscillates around one of the minima of its potential. These oscillations are dumped by the expansion of the Universe, relaxing the initial ALP field value to its vacuum expectation value $\phi_{\rm min}$. The oscillations behave as a classical condensate of ALPs with zero momentum and thus can be the DM constituent as we will show later.

Some of the proposed misalignment mechanisms are the following:

\begin{figure}[t!]
\includegraphics[height=6.5cm]{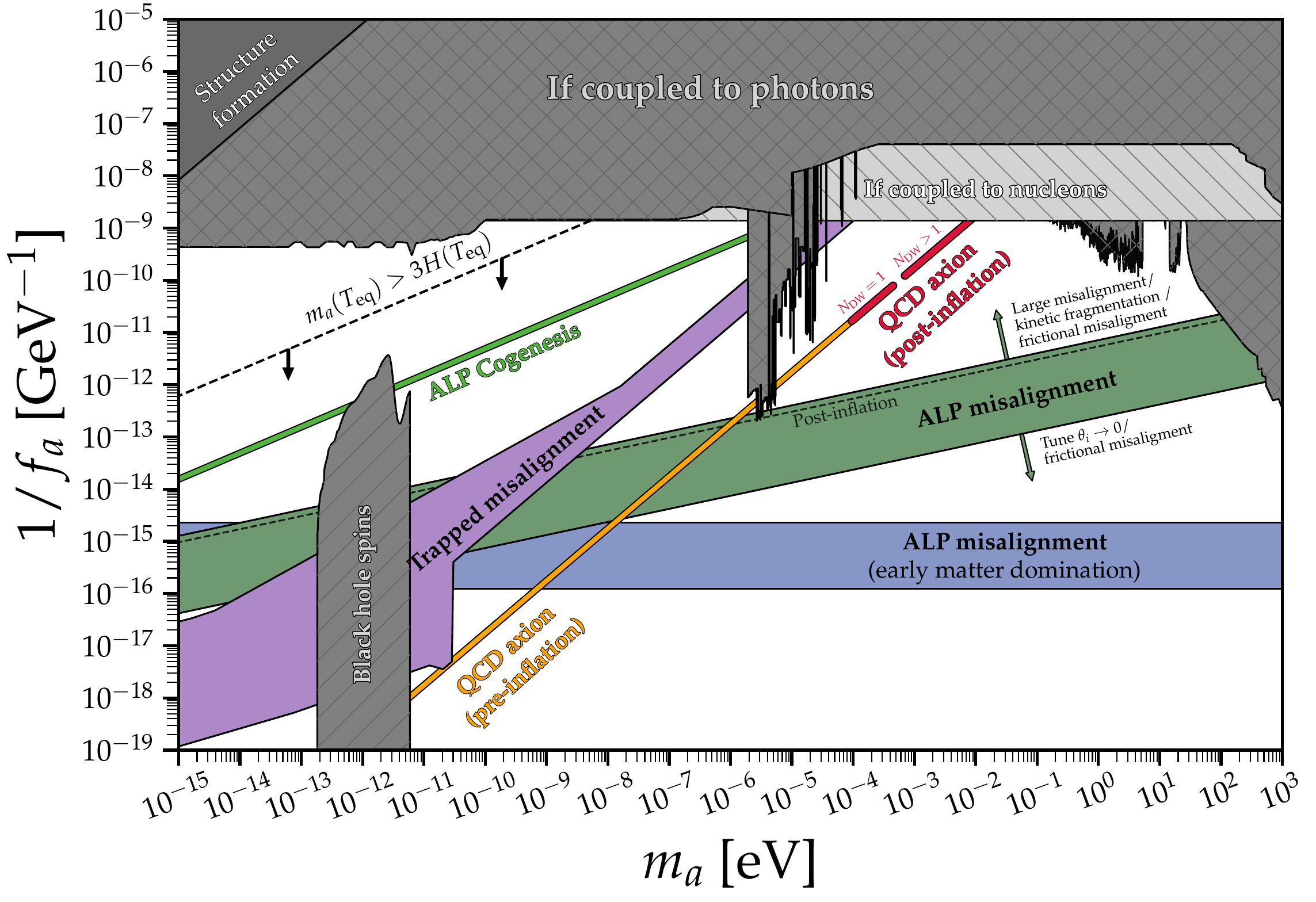}
\centering
\caption{Several proposed misalignment mechanisms and their corresponding parameter space in the $m_\phi-f_\phi$ plane. The plot is taken from \url{https://github.com/cajohare/AxionLimits/tree/master/plots}.} 
\label{fig:misalignment_mech}
\end{figure}

\begin{itemize}
    \item[1.] Conventional misalignment mechanism \cite{Preskill:1982cy, Dine:1982ah,Abbott:1982af}. The ALP is essentially frozen at the initial value $\phi_i$ due to the large Hubble friction effectively behaving as dark energy (DE) with equation of state parameter $\omega_{\phi}=-1$. During radiation domination epoch, when the ALP mass becomes comparable to the Hubble expansion rate at the time $t_{\rm osc}$, $m_\phi\approx 3H(t_{\rm osc})$, the ALP starts oscillating around one of the minima of the potential. For an ultralight QCD axion, it has been argued that the initial misalignment angle required to reproduced the current DM relic density is too small $\theta_i=\phi_i/f_{\phi}\sim \mathcal{O}\left(10^{-3}-10^{-4}\right)$, leading to a fine-tuning problem. However, in Refs.~\cite{Co:2018phi,Co:2018mho} the authors propose a mechanism that allows to choose initial values closed to zero or to $\pi$ without fine-tuning. For example, Higgs inflation dynamics can ensure a large initial axion mass and thus stochastically the axion is relaxed to the minimum of the potential even during inflation. Together with a CP phase shift due to an approximate CP symmetry and a sign flip, the axion can be set to the bottom or the hilltop of the potential, and thus opening the window of light and heavy axion DM. This also helps to relax the isocurvature constraints afflicting the pre-inflationary axion.  
    
    \item[2.] Kinetic misalignment \cite{Co:2019jts,Barman:2021rdr}. In this case, the ALP has large initial velocity in field space, $\dot{\phi}_0$. If the kinetic energy associated to this field velocity is larger than the  potential energy at the time $t_{\rm osc}$, the ALP simply overcomes the potential barriers several times until the kinetic energy redshifts and the ALP gets trapped by the potential. During the stage of large velocity, the ALP behaves as a perfect fluid with $\omega_{\phi}=1$. As a consequence, the oscillations are delayed with respect to the previous scenario. If the initial ALP velocity is not sufficient to jump over a single potential barrier, this mechanism reduces to the conventional misalignment mechanism.
    \item[3.] Trapped scenario \cite{Nakagawa:2020zjr,DiLuzio:2021gos,Kitajima:2023pby}. In this scenario, two different potential minima develop during the Universe evolution from high to low temperatures. Oscillations initially start around a false vacuum $\phi=\phi_*$, due to the temperature dependent potential, allowing $m_{\phi_*}(T)=d^2V(\phi)/d\phi^2 \gtrsim H(T)$ at $\phi_*$. Later on, the true minimum ($\phi=\phi_{\rm min}$) develops, the ALP mass becomes the vacuum mass $m_{\phi_{\rm min}}$, and the false vacuum becomes a maximum. Oscillations around the true minimum start whenever the ALP kinetic energy becomes smaller than the height of the potential barrier and the condition $m_{\phi_{\rm min}}>3H$ is fulfilled. These two stages are thus separated by a strongly non-adiabatic modification of the ALP potential.
    \item[4.] Frictional misalignment \cite{Papageorgiou:2022prc}.  In this case there is an sphaleron-induced thermal friction that effectively contributes to the Hubble friction $H$. Thermal friction can either enhance the axion relic density by delaying the onset of oscillations or suppress it by damping them.
\end{itemize}
Their corresponding region in the $m_\phi-f_\phi$ plane is shown in Fig.~\ref{fig:misalignment_mech}. Compared to conventional misalignment, in the rest of misalignment mechanisms the onset of the oscillations is delayed. However, once the oscillations have begun, the ALP behaves as matter in all the scenarios. In this work, we propose to study the cosmological implications of such an oscillating ALP by generalizing its dynamics prior to the onset of oscillations, treating it as a perfect fluid with a constant equation of state parameter $\omega_{\phi}$. In this work we will preserve the standard cosmological timeline, with a radiation-dominated era followed by matter domination. This will translate into constrains on the model parameters, to ensure that the ALP does not dominate the energy content at early times in the case in which $\omega_{\phi}>1/3$ (see Sec.~\ref{Bayesiananalysis}).
As the ALP serves as the dark matter candidate, its oscillations must commence well within the radiation-dominated epoch, namely, $a_{\rm osc}<10^{-4}$ where $a_{\rm osc}=a(t_{\rm osc})$.
 
\subsection{Cosmological tensions}

In this section, we briefly review the cosmological tensions relevant for our discussion of the ALP parameter space, namely the Hubble and $S_8$ tensions.

The $H_0$ tension refers to the discrepancy between the value of the Hubble constant inferred from CMB observations using the $\Lambda\rm CDM$ model, $H_0 = 67.4 \pm 0.5$ km/s/Mpc \cite{Planck:2018vyg}, and local measurements based on the distance ladder, such as the SH0ES result  $H_0 = 73.04 \pm 1.04$ km/s/Mpc \cite{Riess:2021jrx}. While the former probes the early Universe through the acoustic scale imprinted in the CMB, the latter relies on late-time observations using calibrated standard candles, such as Type Ia supernovae (SNe Ia). This discrepancy may point to new physics beyond $\Lambda\rm CDM$, affecting the expansion history of the Universe.

The $S_8$ tension arises from the mismatch between the amplitude of matter fluctuations inferred from the CMB and that measured by large-scale structure probes. Weak lensing and galaxy clustering surveys typically find lower values, $S_8 \sim 0.76\text{--}0.78$ \cite{KiDS:2020suj,DES:2021wwk,Ivanov:2019pdj,Philcox:2021kcw}, compared to the Planck result $S_8 = 0.834 \pm 0.016$ \cite{Planck:2018vyg}.

A common feature of these tensions is that they can be alleviated by effectively reducing the amount of DM at late times (i.e., after recombination, so as not to affect CMB observables) thereby modifying both the expansion history and the growth of structure. This motivates scenarios in which the ALP is not completely stable, but instead decays into a dark radiation sector, as we develop in the next section.

\section{Pre-inflationary ALP decaying to dark radiation}\label{DRproduction}

\begin{figure}[h]
    \centering
   \subfigure[]{\includegraphics[height=6cm]{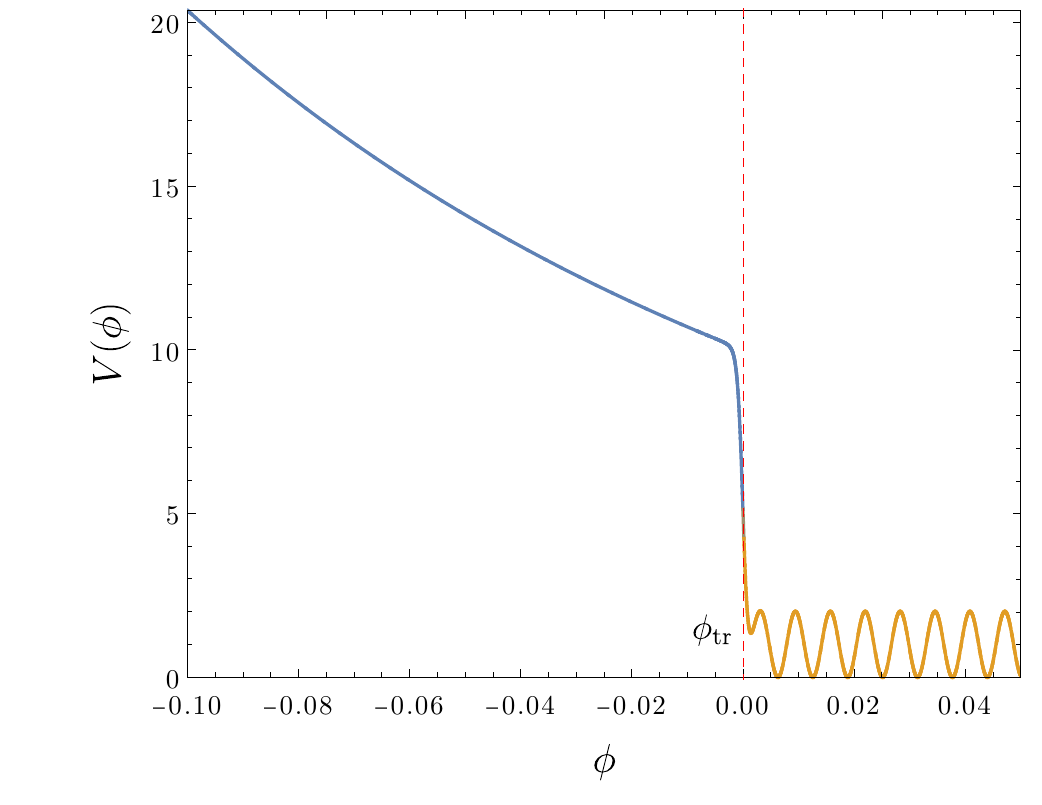}}\hfill
\subfigure[]{\includegraphics[height=6cm]{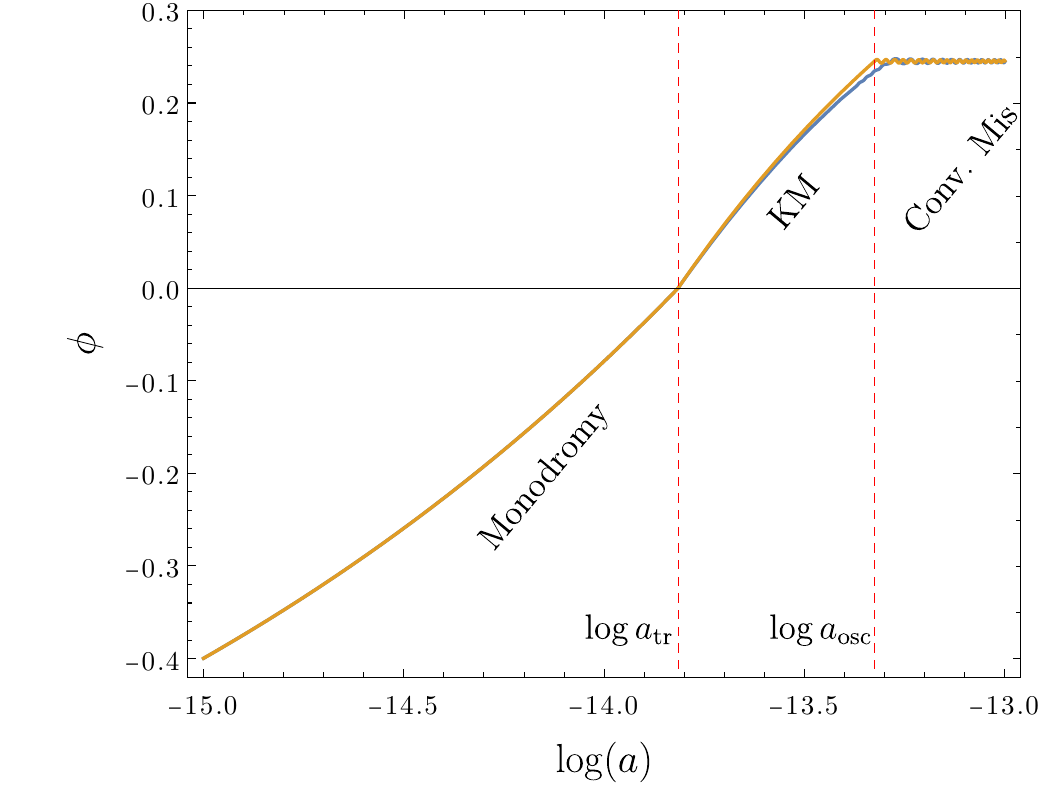}}
    \caption{{\it{Left}}: ALP potential as shown in Eqs.~(\ref{potential}), (\ref{cosinepotential}) and (\ref{potentialofphi}). {\it{Right}}: Evolution of the ALP field across the different regimes. The blue curve shows the exact solution of Eq.~(\ref{alpeom}) for the potential in panel (a), while the orange curve shows the approximate solution. The close agreement between the two  curves justifies the approximations we employed.}\label{fig:potential}
\end{figure}

In this work, we consider a model where a classical ALP condensate experiences a pre-oscillating phase of a constant equation of state $\omega_{\phi}=P_{\phi}/\rho_\phi$ during radiation domination epoch, induced by a particular form of the potential, as will be derived in Subsec.~\ref{sec:background_alp_eom}. However, this stage must eventually end to allow for an oscillatory phase, with the potential settling to a minimum at vanishing potential energy in order for the ALP to reproduce the observed dark matter relic density. One possible realization is to introduce a heavy scalar field, as motivated for instance by string or M-theory or higher-dimensional $\mathcal N=2$ supergravity (see Ref.~\cite{Bertolami:2012xn} and references therein). In this setup, either the field acquires a large vacuum expectation value or it evolves rapidly from an initially frozen state under Hubble friction towards a large field value (e.g., due to a negative exponential potential as the dilaton in Refs.~\cite{Barreiro:1998aj,Brax:2010gi}); in both cases, its dynamics exponentially suppress the part of the potential responsible for a negative equation of state, triggering a transition to a regime where an oscillatory potential dominates. 
Nevertheless, being agnostic about the particular mechanism, we model this transition as
\begin{align}\label{potential}
    V(\phi)=\frac{\exp\left[-2\left(\phi-\phi_{\rm tr}\right)/f_\phi\right]V_\omega(\phi)+V_{\rm osc}(\phi)}{1+\exp\left[-2\left(\phi-\phi_{\rm tr}\right)/f_\phi\right]},
\end{align}
where $V_{\omega}(\phi)$ is the potential that ensures a constant equation of state, $V_{\rm osc}(\phi)$ is the usual cosine-type potential
\begin{align}\label{cosinepotential}
    V_{\rm osc}(\phi)=m_\phi^2f_\phi^2\left[1-\cos\left(\frac{\phi}{f_\phi}\right)\right]
\end{align}
and $\phi_{\rm tr}=\phi(a_{\rm tr})$ is the field value at the scale factor $a_{\rm tr}$, when the phase with constant $\omega_\phi$ ends. Notice that in the limit $\phi-\phi_{\rm tr}<<0$, the exponential dominates in both numerator and denominator and the potential behaves as $V_{\omega}(\phi)$ while in the limit $\phi-\phi_{\rm tr}>>0$, the exponential suppresses the $V_{\omega}(\phi)$ term and the potential behaves as $V_{\rm osc}(\phi)$ and the ALP may eventually oscillate around a minimum at zero potential energy, originating DM as in the standard misalignment scenario. The transition between those regimes is smooth as shown in Fig.~\ref{fig:potential}.

Potentials with regime transitions are commonly used in the literature, such as the Starobinsky potential for the inflaton with singularities \cite{Starobinsky:1992ts} and the models in \cite{Adams:2001vc,Covi:2006ci,Cai:2014vua} that also show a step in the potential shape. 

This ALP is coupled to a dark radiation (DR) sector such that the ALP coherent oscillations generate DR when the decay width $\Gamma_{\phi}$ is larger than the Hubble expansion rate $H$, allowing for the dissipation of part of the DM. The background dynamics of this model may resemble that of the decaying cold dark matter scenario~\cite{Poulin:2016nat,Chudaykin:2016yfk,Chudaykin:2017ptd,Nygaard:2020sow}. An alternative modification of this scenario is presented in Ref.~\cite{Bringmann:2018jpr}, where a phenomenological formula for the DM evolution is proposed. Its cosmological implications are examined in Refs.~\cite{Bringmann:2018jpr,DES:2020mpv,McCarthy:2022gok}. With respect to our work, there are significant differences at the background and perturbation levels. At the background level, in our model, the ALP equation of state before the transition is an input parameter $-1<\omega_{\phi}<1$. This allows the ALP to behave as dark energy ($\omega_{\phi}=-1$) as in conventional misalignment, matter ($\omega_{\phi}=0$), or kination ($\omega_{\phi}=1$) as in kinetic misalignment, prior to the end of the phase with constant $\omega_\phi$. Later on,
the ALP eventually oscillates and decays to DR with a constant decay rate $\Gamma_{\phi}$.
  
In the following, we provide a detailed presentation of the main equations and develop the formalism that will be employed to compute the energy density of ALPs and DR, both at the background and at the perturbation level.

\subsection{Background cosmology}
We assume that the background Universe is well described by the spatially flat FLRW metric
\eq{
ds^2 = g_{\mu\nu}dx^\mu dx^\nu = dt^2-a^2(t)d\vec{x}^2,
}
in comoving coordinates. The derivative of a magntiude with respect to cosmic time will be denoted as $\dot m$. The non vanishing Christoffel symbols in cosmic time are given by
\begin{align*}
    &\Gamma\elev t_{ij} = a\elev 2 H\delta_{ij} \\
    &\Gamma\elev i_{jt}=H\delta\elev i_j
\end{align*}
where $H=\dot a/a$ is the Hubble expansion rate. The two Friedmann equations are
\begin{align}
    &H\elev2=\frac{\rho}{3M_p\elev2}\\
    &\dot H=-\frac{\rho+P}{2M_p\elev2},
\end{align}
where $M_p\elev2=(8\pi G_N)\elev{-1}$ and $G_N$ is the Newtonian gravitation constant.

Sometimes we will use conformal time, defined as follows
\eq{\label{conftime}
dt\equiv a(t)d\eta
}
such that the FLRW metric takes the form
\eq{\label{FRWconformal}
ds^2=a^2(\eta)\left(d\eta^2-d\Vec{x}^2\right).
}
We will use $m'$ to denote the derivative of a magnitude with respect to conformal time. The non-vanishing Christoffel symbols in conformal time are given by
\begin{align}\label{christoffelbackgroundconformal}
    &\Gamma\elev\eta_{\eta\eta}=\mathcal H\\
    &\Gamma\elev\eta_{ij}=\mathcal H\delta_{ij}\\
    &\Gamma\elev i_{j\eta}=\mathcal H\delta\elev i_j,
\end{align}
where $\mathcal H=a'/a$ is the Hubble expansion rate in conformal time. The two Friedmann equations are
\begin{align}\label{friedmanneqconformal}
    &\mathcal H\elev2=a\elev2\frac{\rho}{3M_p\elev2}\\
    &\mathcal H'=-\frac{1}{2}a\elev2\frac{\rho+3p}{3M_p\elev2}
\end{align}

In this work we assume the standard cosmological history of the Universe with the inclusion of the homogeneous ALP field that will be our DM candidate, as we will justify in the next subsection. According to CMB data, DM is present in the Universe previous to the release of the CMB, $z\approx 1100$, when the Universe was still in radiation domination epoch. Hence, using the Friedmann equation, the Hubble expansion rate $H$ as a function of the scale factor during radiation domination epoch is given by
\eq{\label{Hraddom}
H=H_0\sqrt{\Omega^0_{\rm rad}}a^{-2}
}
where $H_0$ is the Hubble expansion rate today, $\Omega^0_{\rm rad}=\rho_{\rm rad}^0/(3H_0^2M_p^2)$ with $\rho^0_{\rm rad}$ the energy density of radiation today and we used that radiation dilutes as $\rho_{\rm rad}\sim a\elev{-4}$. Integrating the above equation to obtain the relation between the scale factor and cosmic time, we get
\begin{align}\label{tinraddom}
    t=\frac{a^2}{2H_0\sqrt{\Omega^0_{\rm rad}}}
\end{align}
so that the Hubble expansion rate can be expressed as
\begin{align}\label{Hubble}
    H=\frac{1}{2t},
\end{align}
valid for a radiation dominated Universe. Analogously, for a matter dominated Universe 
\begin{align}\label{Hmatdom}
    H=H_0\sqrt{\Omega^0_{\rm mat}}a^{-\frac{3}{2}},
\end{align}
with $\Omega_{\rm mat}^0=\rho^0_{\rm mat}/(3H_0^2M_p^2)$, $\rho^0_{\rm mat}$ the energy density of matter today and we used $\rho_m\sim a\elev{-3}$. The relation between time and scale factor is thus given by
\begin{align}\label{tinmatdom}
    t=t_1+\frac{2}{3}\frac{a^{\frac{3}{2}}-a_1^{\frac{3}{2}}}{H_0\sqrt{\Omega^0_{\rm mat}}},
\end{align}
where $a_1$ and $t_1$ is a reference scale factor and time. The Hubble expansion rate as a function of time during matter domination is then
\begin{align}
    H=\frac{2}{3t}.
\end{align}

\subsection{Background ALP equation of motion}\label{sec:background_alp_eom}

Let us study the behavior of the homogeneous ALP field $\phi(t)$ in the expanding Universe. Its equation of motion (eom) comes from the action of an scalar field minimally coupled to gravity
\eq{
S_{\phi}=\int d^4x \sqrt{\lvert g \rvert}\left(\frac{1}{2}\dot{\phi}^2-V(\phi)\right),
}
where $\lvert g\rvert = a^3(t)$ is the determinant of the metric $g_{\mu\nu}$ in comoving coordinates and $V(\phi)$ is the potential we defined in Eq.~(\ref{potential}). This action leads to
\begin{align}\label{alpeom}
    \ddot{\phi}(t)+3H(t)\dot{\phi}(t)+V'(\phi)=0,
\end{align}
and the density and pressure are
\begin{align}
    \rho_\phi=\frac{1}{2}\dot\phi^2+V(\phi),\quad P_\phi=\frac{1}{2}\dot\phi^2-V(\phi).
\end{align}
It only remains to determine the form of the potential during the pre-oscillatory phase that yields a constant equation of state during radiation domination epoch. Following the approach of \cite{vanHolten:2023uja}, we reconstruct the potential that produces a constant ALP equation of state in the limit
 $a<<{a_{\rm tr}}$ and $\phi<<\phi_{\rm tr}$. The requirement of a constant equation of state imposes a relation between the potential and the field velocity
\begin{align}\label{velocity}
 \dot{\phi}=\sqrt{\frac{2\left(1+\omega_{\phi}\right)}{1-\omega_{\phi}}V(\phi)}.   
\end{align}
Multiplying the ALP eom by $\dot{\phi}$ and using the relation between the field velocity and the potential in Eq.~(\ref{velocity}), together with the definition $H=\dot{a}/a$ we obtain
\begin{align}\label{potentialofa}   V(a)=V_{\rm tr}\left(\frac{a_{\rm tr}}{a}\right)^{-3(1+\omega_{\phi})}.
\end{align}
From Eq.~(\ref{velocity}), one can integrate to obtain the evolution of the ALP as a function of the scale factor. Using the expression for the Hubble expansion rate during radiation domination, Eq.~(\ref{Hraddom}), this yields
\begin{align}
    \phi(a)=\begin{cases}
        \phi_{\rm tr}+\frac{2}{H_{\rm tr}}\frac{1}{1-3\omega_{\phi}}\sqrt{\frac{2(1+\omega_{\phi})}{1-\omega_{\phi}}V_{\rm tr}}\left[\left(\frac{a}{a_{\rm tr}}\right)^{\frac{1-3\omega_{\phi}}{2}}-1\right] & \omega_{\phi}\neq\frac{1}{3}\\
        \phi_{\rm tr}+\frac{2\sqrt{V_{\rm tr}}}{H_{\rm tr}}\log\left(\frac{a}{a_{\rm tr}}\right) & \omega_{\phi}=\frac{1}{3}
    \end{cases}
\end{align}
where $H_{\rm tr}=H_0\sqrt{\Omega^0_{\rm rad}}a_{\rm tr}^{-2}$ is the Hubble expansion rate at the transition. The desired potential is then obtained by inverting the above expressions and substituting the result into Eq.~(\ref{potentialofa})
\begin{align}\label{potentialofphi} V(\phi)=\begin{cases}    V_{\rm tr}\left[\frac{1}{2}(\phi-\phi_{\rm tr})(1-3\omega_{\phi}) H_{\rm tr}\sqrt{\frac{1-\omega_{\phi}}{2(1+\omega_{\phi})V_{\rm tr}}}+1\right]^{\frac{-6(1+\omega_{\phi})}{1-3\omega_{\phi}}} & \omega_{\phi}\neq \frac{1}{3}\\
V_{\rm tr}\exp\left[(\phi-\phi_{\rm tr})\frac{2 H_{\rm tr}}{\sqrt{V_{\rm tr}}}\right] & \omega_{\phi}=\frac{1}{3}.
    \end{cases}
\end{align}
For $\omega_\phi\leq 1/3$, the power is negative and these potentials belong to the class of {\it{tracking}} potentials \cite{Ratra:1987rm,Steinhardt:1999nw,Sahni:1999gb,Tsujikawa:2013fta}, for which the scalar field dynamics asymptotically converge toward an attractor solution, rendering the evolution insensitive to the choice of initial conditions. This property is no longer true for $\omega_\phi>1/3$, and the initial conditions must be adjusted for the field to realize the desired value of $\omega_\phi$. Nevertheless, we will include this case in our analysis. The constant $V_{\rm tr}$, corresponding to the value of the ALP potential during the transition, is undetermined. Using Eq.~(\ref{velocity}), the energy density of the ALP during the phase with constant $\omega_\phi$ is given by
\begin{align}
    \rho_{\phi}(a)=\frac{1}{2}\dot\phi^2+V=\frac{2}{1-\omega_{\phi}}V(a),
\end{align}
and at the transition
\begin{align}
    \rho_{\phi}(a_{\rm tr})=\frac{2}{1-\omega_{\phi}}V_{\rm tr}.
\end{align}
Let us define the parameter
\begin{align}
    r_{\phi}=\frac{\rho_\phi(a_{\rm tr})}{2m_\phi^2f_\phi^2}=\frac{1}{1-\omega_{\phi}}\frac{V_{\rm tr}}{m_\phi^2f_\phi^2}
\end{align}
that measures the ratio between the ALP energy density at the transition and the height of the cosine potential barrier.
Depending on the value of the parameter $r_\phi$, we can distinguish three different scenarios just after the transition:
\begin{itemize}
    \item [1.] $r_{\phi}<1$. In this case the ALP does not have enough energy density to overcome the cosine potential barrier and oscillates around a minimum at non-vanishing potential energy, behaving as DE, $\rho_\phi\sim \rm const$.
    \item [2.] $r_{\phi}\simeq 1$. In this case the ALP has enough energy to overcome the barrier that separates the minimum at non zero potential energy from the adjacent minimum at zero potential energy. The oscillatory phase then begins approximately at the top of the cosine potential provided that the condition $m_\phi>3H$ is fulfilled~\cite{Co:2019jts}. During this phase the ALP energy density redshifts as matter, $\rho_\phi\sim a^{-3}$ as we show below. 
    \item [3.] $r_{\phi}\gtrsim 1$. In this case, the available energy is enough to allow the ALP to explore several potential minima, entering in a phase of kinetic misalignment. The field velocity at the transition is given by   
    \begin{align}
        \dot{\phi}_{\rm tr}=2m_\phi f_\phi\sqrt{r_{\phi}}
    \end{align}    
  During the kinetic misalignment phase, the field velocity evolves as  $\dot{\phi} \propto a^{-3}$ (obtained by setting $V=0$ in Eq.~(\ref{alpeom})), implying $\rho_\phi \propto a^{-6}$.
By evolving this regime until $\rho_\phi = 2Cm_\phi^2 f_\phi^2$, with $C$ a numerical coefficient of order $C \simeq 1/2$, defines $a_{\rm osc}$, the scale factor at which the ALP energy density redshifts down to the height of the cosine potential barrier
\begin{align}\label{a2}
    \frac{1}{2}\dot{\phi}^2_{\rm tr}\left(\frac{a_{\rm tr}}{a_{\rm osc}}\right)^6=2m_\phi^2f_\phi^2\rightarrow a_{\rm osc}=a_{\rm tr}\left(\frac{r_{\phi}}{C}\right)^{\frac{1}{6}}.
\end{align}

 For $ a>a_{\rm osc} > a_{\rm tr}$, the ALP no longer possesses sufficient energy to overcome the potential barrier, and the oscillatory phase begins typically near the top of the cosine potential if $m_\phi > 3H$, with $\rho_\phi \propto a^{-3}$, as in the previous case. Notice that in case (2), we have $a_{\rm osc}\approx 2^{1/6} a_{\rm tr} \approx a_{\rm tr}$.
\end{itemize}
In this work we will ignore the first case since the ALP is our DM candidate. Besides, we will always assume that when the ALP gets trapped in a minimum of vanishing potential energy, the condition $m_\phi>3H$ is fulfilled. The oscillatory phase must begin during radiation domination epoch so that the ALP can behave as matter and dominate over radiation at later times, around recombination. 

Now we show that the ALP energy density effectively behaves as matter during the oscillatory phase when $a>a_{\rm osc}$. From the ALP eom in Eq.~(\ref{alpeom}) using the potential in Eq.~(\ref{cosinepotential})
we get\footnote{Here we ignore the ALP coupling to DR. Its effect will be later included by using energy conservation (see Subsec.~\ref{subsec:pert_cosmology})}
\eq{
\ddot{\phi}+3H\dot{\phi}+m_{\phi}^2f_\phi\sin\left(\frac{\phi}{f_\phi}\right)=0.
}
Around a minimum, we redefine $\phi=\phi_{\rm min}+\varphi$ and approximate $\sin(\varphi/f_\phi)\approx\varphi/f_\phi$ so that
\eq{
\ddot{\varphi}+3H\dot{\varphi}+m_{\phi}^2\varphi=0.
}
The solution to this equation during the radiation domination epoch is given by a linear combination of Bessel functions
\eq{
\label{alpposition}
&\varphi=
a^{-\frac{1}{2}}\left[b_1 J_{1/4}\left(\frac{m_{\phi}}{2H}\right)+b_2 Y_{1/4}\left(\frac{m_{\phi}}{2H}\right)\right]\\
\label{alpvelocity}&
\dot{\varphi}=-m_{\phi}a^{-\frac{1}{2}}\left[b_1 J_{5/4}\left(\frac{m_{\phi}}{2H}\right)+b_2 Y_{5/4}\left(\frac{m_{\phi}}{2H}\right)\right]
}
where $b_1$ and $b_2$ are integration constants ensuring the continuity of the field and velocity in the regimen transition, $J_{1/4}(x), J_{5/4}(x)$ are Bessel functions of first kind or order $1/4$ and $5/4$, respectively, while $Y_{1/4}(x), Y_{5/4}(x)$ are of second kind and order $1/4$ and $5/4$. In this work we assume that at $a=a_{\rm osc}$ the usual condition $m_{\phi}>3H$ is satisfied so that the ALP does not get frozen and effectively starts oscillating. In this situation, the above expressions can be simplified 
\begin{align}
&\varphi=\varphi_{\rm osc} \left(\frac{a_{\rm osc}}{a}
\right)^{\frac{3}{2}}\cos\left(m_\phi t+\nu\right)\\
&\dot{\varphi}=-\varphi_{\rm osc}\left(\frac{a_{\rm osc}}{a}\right)m_{\phi}\sin\left(m_\phi t+\nu\right)
\end{align}
where $\varphi_{\rm osc}$ and $\nu$ are integration constants and we neglected terms of order $H/m_\phi$. The limit of the Bessel functions for large argument is independent of the background cosmology.

The energy density of the ALP field during oscillations is thus
\eq{\label{ALPenergydensity}
\rho_\phi(a)\approx\frac{1}{2}\dot{\varphi}^2+\frac{1}{2}m_\phi^2\varphi^2 = \frac{1}{2}m_\phi^2\varphi_{\rm osc}^2\left(\frac{a_{\rm osc}}{a}\right)^3.
}
Notice that after the onset of the oscillations the ALP energy density scales as $\rho_\phi\sim a^{-3}$, independently of the background cosmology since we neglected $H/m_{\phi}$. Analogously, the pressure is given by
\eq{
P_\phi(a)\approx \frac{1}{2}\dot{\varphi}^2-\frac{1}{2}m_\phi^2\varphi^2 =-\frac{1}{2}m_\phi^2\varphi_{\rm osc}^2\left(\frac{a_{\rm osc}}{a}\right)^3\cos\left[2\left(m_\phi t  + \nu\right)\right],
}
where we used Eq.~(\ref{Hubble}) to write the pressure as a function of time. In a period $T=\pi/m_\phi$ the scale factor changes an amount
\eq{\label{change_scale_factor}
\Delta a\approx \dot{a}\Delta t = H a \frac{\pi}{m_\phi}\rightarrow \frac{\Delta a}{a}=\pi\frac{H}{m_\phi},
}
that is negligible after the onset of the oscillations. Hence, averaging over a period we have $\langle \cos\left[2(m_\phi t + \nu)\right] \rangle_T = 0$ such that the pressure is
$P_\phi(a) = 0$. Hence the ALP behaves as cold matter when oscillations have begun, justifying that the ALP can be the DM that will dominate the energy content of the Universe after recombination. On the other hand, since oscillations typically begin on top of the cosine potential, we will take $\varphi_{\rm osc}\approx\pi f_{\phi}$ and ignore anharmonicity effects. Sufficient DM will be produced if oscillations begin well inside radiation domination epoch.

\subsection{Background dark radiation production by ALPs}

\subsubsection{Occupation number}
In this work we will assume that the ALP possesses a trilinear coupling to a dark radiation sector, namely
\eq{
\lag_{\phi\rm DR}=-\frac{1}{8}g_{\phi\rm DR}\phi\frac{\epsilon^{\mu\nu\rho\lambda}}{\sqrt{\lvert g \rvert}}F_{\mu\nu}F_{\rho\lambda},
}
where $\epsilon^{\mu\nu\rho\lambda}$ is the totally antisymmetric Levi-Civita tensor with $\epsilon^{0123}=+1$, $g_{\phi\rm DR}$ is a dimensionful coupling with energy units of $E^{-1}$ and $\phi=\phi_{\rm min}+\varphi$. Perturbativity imposes $g_{\phi\rm DR}<4\pi/f_\phi$. DR is sourced by ALP oscillations and its production  becomes effective when $\Gamma_\phi>H$, where $\Gamma_\phi$ is the ALP decay width, which we compute below.

The action of dark photons coupled to the ALP is
\eq{
S_{\rm DR}=\int d^4x \sqrt{\lvert g\rvert}\left(-\frac{1}{4}F_{\mu\nu}F^{\mu\nu}+\lag_{\phi\rm DR}\right),
}
where $F_{\mu\nu}$ is the dark photon strength tensor $F_{\mu\nu}=\partial_\mu A_\nu-\partial_\nu A_\mu$. Hence, from the point of view of the DR, the ALP is just a background field that continuously supports particle production. At leading order, we consider that the ALP simply follows the eom (\ref{alpeom}) with solution Eqs.~(\ref{alpposition}), (\ref{alpvelocity}), and thus we will neglect the backreaction of the dark photons on the ALP field.
This effect will be later included when we impose energy conservation.

Using temporal gauge $A_0=0$ and conformal time defined in Eq.~(\ref{conftime}) with $\sqrt{\lvert g\rvert}=a^4$, and integrating by parts this reduces to
\eq{\label{photonaction}
S_{\rm DR}\equiv \int d^4x~ \lag_{\rm DR}=\int d^4x \left(\frac{1}{2}A_i'^2-\frac{1}{2}\left[\left(\partial_i A_j\right)^2-\left(\partial_i A_i\right)^2\right]+\frac{1}{2}g_{\phi\rm DR}\varphi'\epsilon^{ijk}A_i\partial_jA_k\right)
}
where $i,j,k=1-3$ are spatial indices. In order to quantize the theory, we write $A_i$ in terms of creation and annihilation operators
\eq{\label{quantizedphoton}
A_i=\int \frac{d^3k}{(2\pi)^3}\sum_{\lambda=\pm}\left(\epsilon_i(\Vec{k},\lambda)\mu_{k\lambda}(\eta)a_{\Vec{k}\lambda}e^{i\Vec{k}\Vec{x}}+\epsilon^{*}_i(\Vec{k},\lambda)\mu^{*}_{k\lambda}(\eta)a^{\dagger}_{\Vec{k}\lambda}e^{-i\Vec{k}\Vec{x}}\right),
}
where the mode functions $\mu_{k\lambda}$ only depend on the modulus of $\lvert\vec{k}\rvert=k$ due to the isotropy of the FLRW metric and $\lambda=\pm$ are the two polarizations of the photon. Creation and annihilation operator satisfy canonical commutation relations
\eq{\label{creationandannihilationoperators}
[a_{\Vec{k}\lambda},a^{\dagger}_{\Vec{k'}\lambda'}]=(2\pi)^3\delta^3(\Vec{k}-\Vec{k}')\delta_{\lambda\lambda'}.
}
Notice that the delta function is related to the space volume through
\eq{
\lim_{\vec{k}\rightarrow \vec{k}'}\delta^{(3)}(\vec{k}-\vec{k}')=\lim_{\vec{k}\rightarrow \vec{k}'}\int d^3x~e^{i(\vec{k}-\vec{k}')\vec{x}}= V(\rightarrow \infty),
}
and thus having energy units of $E^{-3}$. This implies that the operators creation and annihilation operators $a^{\dagger}_{k,\lambda}$, $a_{k\lambda}$ have dimension $E^{-\frac{3}{2}}$. Consequently, since $A_{i}$ has dimension of $E$, the mode functions have dimension $E^{-\frac{1}{2}}$. 

On the other hand, the circular polarization vectors obey
\eq{\label{polarizationvectorrelation}
\epsilon(\vec{k},\lambda)\cdot\epsilon^{*}(\vec{k},\lambda')=\delta_{\lambda \lambda'},\quad \vec{k}\cdot \epsilon(\vec{k},\lambda)=0,\quad \vec{k}\times \epsilon(\vec{k},\lambda)=-i k \lambda \epsilon(\vec{k},\lambda), \quad \epsilon(-\vec{k},\lambda)=\epsilon^*(\vec{k},\lambda)
}
where the last two relations come from the fact that the two linear polarization vectors and the direction of propagation $\vec{k}$ constitute a direct trihedral.

The classical eom of the photon field is thus given by
\eq{
A_i''-\partial_j^2 A_i+\delta_{ij}\partial_j^2 A_i-g_{\phi\rm DR}\varphi'\epsilon^{ijk}\partial_j A_k=0.
}
Using the relations in Eq.~(\ref{polarizationvectorrelation}) the mode functions satisfy
\eq{\label{modefunctioneom}
\mu_{k\lambda}''+(k^2-\lambda g_{\phi\rm DR} k\varphi')\mu_{k\lambda}=0,
}
which is the eom of a harmonic oscillator with a time-dependent frequency given by
\eq{\label{omega}
\omega^2_{k\lambda}=k^2-k\lambda g_{\phi\rm DR} \varphi'.
}
In the following, we will derive physical arguments to fix the two initial conditions for the mode functions.

The first initial condition comes from imposing canonical commutation relations between the dark photon field $A_i$ and its corresponding conjugated momentum $\Pi_i$. This can be obtained trivially from the action in Eq.~(\ref{photonaction})

\eq{\label{conjugatedmomentum}
\Pi_i=\frac{\partial\lag_\gamma}{\partial A_i'}=A_i'=\int \frac{d^3k}{(2\pi)^3}\sum_{\lambda=\pm}\left(\epsilon_i(\Vec{k},\lambda)\mu'_{k\lambda}(\eta)a_{\Vec{k}\lambda}e^{i\Vec{k}\Vec{x}}+\epsilon^{*}_i(\Vec{k},\lambda)\mu'^{*}_{k\lambda}(\eta)a^{\dagger}_{\Vec{k}\lambda}e^{-i\Vec{k}\Vec{x}}\right).
}
The canonical commutation relation at equal times imposes
\eq{\label{canonicalcommutator}
\left[A_i(\eta,\vec{x}),\Pi_j(\eta,\vec{y})\right]\equiv i\delta_{ij}\delta^{(3)}_{\rm tr}(\vec{x}-\vec{y}),
}
where the \textit{transverse} delta is defined through
\eq{
\delta_{ij}\delta^{(3)}_{\rm tr}(\vec{x}-\vec{y})\equiv\int\frac{d^3k}{(2\pi)^3}e^{i\vec{k}(\vec{x}-\vec{y})}\left(\delta_{ij}-\frac{k_ik_j}{k^2}\right).
}
Using the expansion of the dark photon field in Eq.~(\ref{quantizedphoton}) and the relations in Eqs.~(\ref{creationandannihilationoperators}) and (\ref{polarizationvectorrelation}), we arrive to
\eq{
\left[A_i(\eta,\vec{x}),\Pi_j(\eta,\vec{y})\right]=\int\frac{d^3k}{(2\pi)^3}\sum_{\lambda=\pm}\left(\epsilon_i(\vec{k},\lambda)\epsilon^*_j(\vec{k},\lambda)\mu_{k\lambda}\mu'^*_{k\lambda}e^{i\vec{k}\cdot(\vec{x}-\vec{y})}-\epsilon_i(-\vec{k},\lambda)\epsilon^*_j(-\vec{k},\lambda)\mu^*_{k\lambda}\mu'_{k\lambda}e^{-i\vec{k}\cdot(\vec{x}-\vec{y})}\right).
}
For the second term we change $\vec{k}\rightarrow-\vec{k}$, such that
\eq{
\left[A_i(\eta,\vec{x}),\Pi_j(\eta,\vec{y})\right]=\int\frac{d^3k}{(2\pi)^3}e^{i\vec{k}\cdot(\vec{x}-\vec{y})}\sum_{\lambda=\pm}\epsilon_i(\vec{k},\lambda)\epsilon^*_j(\vec{k},\lambda)\left(\mu_{k\lambda}\mu'^*_{k\lambda}-\mu^*_{k\lambda}\mu'_{k\lambda}\right).
}
It is easy to check that this quantity is time independent because the term in parenthesis is the Wroskian, $W(\mu_{k\lambda},\mu^*_{k\lambda})$
\eq{
&W(\mu_{k\lambda},\mu^*_{k\lambda})\equiv
\mu_{k\lambda}\mu'^*_{k\lambda}-\mu^*_{k\lambda}\mu'_{k\lambda} \\&W'(\mu_{k\lambda},\mu^*_{k\lambda})=\mu_{k\lambda}\mu''^*_{k\lambda}-\mu^*_{k\lambda}\mu''_{k\lambda}=-\omega^2_{k\lambda}\lvert\mu_{k\lambda}\rvert^2+\omega^2_{k\lambda}\lvert\mu_{k\lambda}\rvert^2=0.
}
Hence, we can compute the commutator at the temporal infinity when $a\gg a_{\rm osc}$ such that $\theta'\rightarrow 0$ and $\omega_{k\lambda}$ does not depend on $\lambda$ (see Eqs.~(\ref{alpvelocity}) and (\ref{omega})) so neither does $\mu_{k\lambda}$. Hence it can be factor out from the sum over polarizations
\begin{eqnarray}\label{result}
 \left[A_i(\vec{x}),\Pi_j(\vec{y})\right]_{\infty}&=&\int\frac{d^3k}{(2\pi)^3}e^{i\vec{k}\cdot(\vec{x}-\vec{y})}W(\mu_{k\lambda}\mu^*_{k\lambda})_{\infty}\sum_{\lambda=\pm}\epsilon_i(\vec{k},\lambda)\epsilon^*_j(\vec{k},\lambda) \nonumber \\
 &=&\int\frac{d^3k}{(2\pi)^3}e^{i\vec{k}\cdot(\vec{x}-\vec{y})}W(\mu_{k\lambda}\mu^*_{k\lambda})_{\infty}\left(\delta_{ij}-\frac{k_ik_j}{k^2}\right),   
\end{eqnarray}
where the subscript $\infty$ means that the quantities are evaluated at the temporal infinity. Thus comparing Eqs.~(\ref{canonicalcommutator}) and (\ref{result}) we obtain
\eq{\label{wroskiancondition}
W(\mu_{k\lambda}\mu^*_{k\lambda})=i,
}
where we used that the Wroskian is time independent such that it can be computed at any time.

The second initial condition comes from considerations about the Hamiltonian. This is calculated as follows
\eq{
\mathcal{H}_\gamma=A'_i\Pi_i-\lag_\gamma=\frac{1}{2}\Pi_i^2+\frac{1}{2}\left[\left(\partial_i A_j\right)^2-\left(\partial_i A_i\right)^2\right]-\frac{1}{2}g_{\phi\rm DR}\varphi'\epsilon^{ijk}A_i\partial_jA_k.
}
Now using the expansion of the photon field and its conjugated momentum in Eqs.~(\ref{quantizedphoton}), (\ref{conjugatedmomentum}), the commutation relations between creation and annihilation operators in Eq.~(\ref{creationandannihilationoperators}) and the properties of the polarization vectors in Eq.~(\ref{polarizationvectorrelation}), we obtain
\eq{\label{hamiltonian}
H=\int d^3x~\mathcal{H}_\gamma &=\frac{1}{2} \int\frac{d^3k}{(2\pi)^3}\sum_{\lambda=\pm}\left(\lvert\mu'_{k\lambda}\rvert^2+\omega^2_{k\lambda}\lvert\mu_{k\lambda}\rvert^2\right)(a^{\dagger}_{\vec{k}\lambda}a_{\vec{k}\lambda}+a_{\vec{k}\lambda}a^{\dagger}_{\vec{k}\lambda})\nn\\
&+\frac{1}{2}\int\frac{d^3k}{(2\pi)^3}\sum_{\lambda=\pm}\left(\mu'^2_{k\lambda}+\omega^2_{k\lambda}\mu^2_{k\lambda}\right)a_{\vec{k}\lambda}a_{-\vec{k}\lambda}+\hc\equiv \int\frac{d^3k}{(2\pi)^3}\sum_{\lambda=\pm} h_{k\lambda},
}
where $h_{k\lambda}$ has units of $E^{-2}$. The Hamiltonian is time-dependent and can only be diagonalized at a fixed time $\eta_0$. This can be done just by imposing the condition\footnote{This condition is always fulfilled in Minkowski space-time without background fields where the mode functions are $\mu_k=\frac{1}{\sqrt{2\omega_k}}e^{-i\omega_k\eta}$ such that the Hamilitonian is diagonal for any time.}
\eq{\label{instantaneousdiagonalization}
\mu'_{k\lambda}(\eta_0)=-i\omega_{k\lambda}(\eta_0)\mu_{k\lambda}(\eta_0),
}
providing the second initial condition for the mode functions. 

Finally, in the Heisenberg picture (the operators evolve in time but the quantum states do not evolve) we start with the vacuum state $\rvert 0\rangle$ (zero particles) at $\eta_0$. Thus the \textit{single particle occupation number} $n_{k\lambda}$ is defined from the Hamiltonian as\footnote{Notice that in QFT with Hamiltonian 
$h_{k}=\frac{1}{2}\omega_k(a^{\dagger}_k a_k+a_k a^{\dagger}_k)$, with $\omega_{k}$ time independent such that there is no cross terms in the Hamiltonian, and multiparticle states $\rvert p_1...p_n\rangle$, the expectation value of the Hamiltonian is
\eq{
\frac{\langle p_1...p_n\lvert h_k\rvert p_1...p_n\rangle}{\langle p_1...p_n\lvert p_1...p_n\rangle}=\omega_{k\lambda}\frac{\langle p_1...p_n\lvert a^{\dagger}_{k\lambda}a_{k\lambda}+\frac{1}{2}V\rvert p_1...p_n \rangle}{\langle p_1...p_n\lvert p_1...p_n\rangle}=\omega_{k\lambda} V\left(\frac{\langle p_1...p_n\lvert a^{\dagger}_{k\lambda}a_{k\lambda}\rvert p_1...p_n \rangle}{V\langle p_1...p_n\lvert p_1...p_n\rangle}+\frac{1}{2}\right)\equiv \omega_{k\lambda} V\left(n_{k\lambda}+\frac{1}{2}\right),\nn
}
where in the last step we have defined the dimensionless quantity
\eq{\nn
n_{k\lambda}=\frac{\langle p_1...p_n\lvert a^{\dagger}_{k\lambda}a_{k\lambda}\rvert p_1...p_n \rangle}{V\langle p_1...p_n\lvert p_1...p_n\rangle},
}
that counts how many particles with momentum $\vec{k}$ per unit volume are there in the corresponding multiparticle state.
Our case is analogous but the responsible of a non-vanishing $n_{k\lambda}$ is the time dependent Hamiltonian, that makes $\rvert 0 \rangle$ to be an eigenstate with zero particles only at $\eta_0$.
}
\eq{\label{singleparticle}
\langle 0\lvert h_{k\lambda} \rvert 0\rangle\equiv\omega_{k\lambda}\left(n_{k\lambda}+\frac{1}{2}\right)V,
}
where $V$ is the (infinite) volume and $n_{k\lambda}$ is the number of particles with momentum between $\vec{k}$ and $\vec{k} + d\vec{k}$ per polarization and per unit volume 
\eq{
n_{k\lambda}=\frac{d n_\lambda}{d^3x d^3k}.
}
This definition tells us that the non diagonal form of the Hamiltonian for times $n>\eta_0$ can be interpreted in terms of creation of particles from the vacuum. Mathematically this is because the state $\rvert 0 \rangle$ is only an eigenstate of the hamiltonian at the time $\eta_0$. On the other hand, we can evaluate the left-hand side to obtain
\eq{
\langle 0\lvert h_{k\lambda} \rvert 0\rangle=\frac{1}{2}\left(\lvert\mu'_{k\lambda}\rvert^2+\omega^2_{k\lambda}\lvert\mu_{k\lambda}\rvert^2\right)V,
}
where we used 
\eq{
\langle 0\lvert a_{k\lambda}a^{\dagger}_{k\lambda}\rvert 0 \rangle = \langle 0\lvert a^{\dagger}_{k\lambda} a_{k\lambda} + V \rvert 0 \rangle = V\langle 0\lvert 0 \rangle = V, \textrm{ with }\langle 0\lvert 0\rangle =1.
}
Hence
\eq{\label{occupationnumber}
n_{k\lambda}=\frac{\lvert\mu'_{k\lambda}\rvert^2+\omega^2_{k\lambda}\lvert\mu_{k\lambda}\rvert^2}{2\omega_{k\lambda}}-\frac{1}{2}.
}
In other words, initially in the vacuum state in Heisenberg picture, there are no particles due to the condition in Eq.~(\ref{instantaneousdiagonalization}). However, the background field creates particles out of the vacuum with energy given by Eq.~(\ref{singleparticle}).

\subsubsection{WKB approach}
In this work we will use the adiabatic WKB representation to solve the eom (\ref{modefunctioneom})
\eq{\label{adianaticrep1}
\mu_{k\lambda}\equiv\frac{\alpha_{k\lambda}(\eta)}{\sqrt{2\omega_{k\lambda}(\eta)}}e^{-i\Psi_{k\lambda}(\eta)}+\frac{\beta_{k\lambda}(\eta)}{\sqrt{2\omega_{k\lambda}(\eta)}}e^{i\Psi_{k\lambda}(\eta)},
}
where $\alpha_{k\lambda}$, $\beta_{k\lambda}$ are the \textit{Bogoliubov} coefficients and $\Psi_{k\lambda}$ is the accumulated phase
\eq{
\Psi_{k\lambda}(\eta)\equiv\int_{\eta_0}^{\eta}d\eta'\omega_{k\lambda}(\eta').
}
The time derivative is taken as if the Bogoliubov coefficients and $\omega_{k\lambda}$ were time independent
\eq{\label{adiabaticrep2}
\mu'_{k\lambda}\equiv-i\alpha_{k\lambda}(\eta)\sqrt{\frac{\omega_{k\lambda}(\eta)}{2}}e^{-i\Psi_{k\lambda}(\eta)}+i\beta_{k\lambda}(\eta)\sqrt{\frac{\omega_{k\lambda}(\eta)}{2}}e^{i\Psi_{k\lambda}(\eta)}.
}
This condition implies non-trivial relations among the Bogoliubov coefficients
\eq{
\alpha'_{k\lambda}(\eta)=\frac{\omega'_{k\lambda}(\eta)}{2\omega_{k\lambda}(\eta)}\beta_{k\lambda}(\eta)e^{2i\Psi_{k\lambda}(\eta)},\quad \beta'_{k\lambda}(\eta)=\frac{\omega'_{k\lambda}(\eta)}{2\omega_{k\lambda}(\eta)}\alpha_{k\lambda}(\eta)e^{-2i\Psi_{k\lambda}(\eta)}.
}
It is easy to check that the WKB mode functions are then solutions of Eq.~(\ref{modefunctioneom}). The Wroskian condition in Eq.~(\ref{wroskiancondition}) imposes the following normalization of the Bogoliubov coefficients
\eq{\label{bogoliubovnormalization}
\lvert \alpha_{k\lambda}\rvert^2-\lvert \beta_{k\lambda}\rvert^2=1.
}
On the other hand, combining the instantaneous diagonalization of the Hamiltonian in Eq.~(\ref{instantaneousdiagonalization}) and the normalization of the Bogoliubov coefficients one arrives to the initial conditions for the Bogoliubov coefficients
\eq{
\alpha_{k\lambda}(\eta_0)=1,\quad \beta_{k\lambda}(\eta_0)=0.
}
On the mode functions this translates into
\eq{\label{modefunctioninitialconditions}
\mu_{k\lambda}(\eta_0)=\frac{1}{\sqrt{2\omega_{k\lambda}(\eta_0)}},\quad \mu'_{k\lambda}(\eta_0)=-i\sqrt{\frac{\omega_{k\lambda}(\eta_0)}{2}}.
}

Using the expression for the single particle occupation number in Eq.~(\ref{occupationnumber}) and the adiabatic representation for the mode functions in Eqs.~(\ref{adianaticrep1}) and (\ref{adiabaticrep2}) we obtain
\eq{
n_{k\lambda}=\frac{\omega_{k\lambda}\lvert\beta_{k\lambda}\rvert^2}{\omega_{k\lambda}}=\lvert\beta_{k\lambda}\rvert^2,
}
meaning that particle production only depends on the Bogoliubov coefficient $\beta_{k\lambda}$ that parametrizes the departure of the mode function from the positive frequency mode. Since we expect small particle occupation numbers, one can do the following approximations
\eq{\label{betaeq}
\alpha_{k\lambda}\approx 1,\quad \beta'_{k\lambda}(\eta)=\frac{\omega'_{k\lambda}(\eta)}{2\omega_{k\lambda}(\eta)}e^{-2i\Psi_{k\lambda}(\eta)},
}
and we only need to compute $\beta_{k\lambda}$.
Eq.~(\ref{betaeq}) will be used to obtain an analytic insight of the dark photon production. 

\subsubsection{Mathieu analysis}
Let us carefully inspect the eom for the mode functions in Eq.~(\ref{modefunctioneom}) when the ALP is already oscillating around one of the minima of its potential for $a>a_{\rm osc}$.
Using Eq.~(\ref{alpvelocity}) and the relation $\varphi'=a\dot\varphi$ we can calculate the frequency 
in Eq.~(\ref{omega}) and get
\eq{
\omega^2_{k\lambda}=k^2-k\lambda g_{\phi\rm DR} \varphi_{\rm osc}a^{\frac{3}{2}}_{\rm osc}m_\phi a^{-\frac{1}{2}}\cos(m_\phi t+\nu).
}
To convert Eq.~(\ref{modefunctioneom}) in a Mathieu equation
\eq{\label{generalmathieu}
\frac{d^2y}{dx^2}+\left[p-2q\cos(2x)\right]y=0
}
we come back to cosmic time obtaining 
\eq{
\ddot{\mu}_{k\lambda}+\dot{\mu}_{k\lambda}H+\frac{k^2-k\lambda g_{\phi\rm DR} \varphi_{\rm osc}a^{\frac{3}{2}}_{\rm osc}m_\phi a^{-\frac{1}{2}}\cos(m_\phi t+\nu)}{a^2}\mu_{k\lambda}=0.
}
Now we define a new variable $z\equiv \frac{1}{2}(m_\phi t + \nu)$, such that
\eq{
\frac{d^2\mu_{k\lambda}}{dz^2}+2\frac{H}{m_\phi}\frac{d\mu_{k\lambda}}{dz}+4\frac{k^2-k\lambda g_{\phi\rm DR} \varphi_{\rm osc}a^{\frac{3}{2}}_{\rm osc}m_\phi a^{-\frac{1}{2}}\cos(2z)}{m_\phi^2 a^2}\mu_{k\lambda}=0.
}
During oscillations we can neglect $H/m_\phi$ and rearrange the last term to finally arrive to the result
\eq{\label{mymathieu}
\frac{d^2\mu_{k\lambda}}{dz^2}+\left[\frac{4k^2}{m_\phi^2 a^2}-2\lambda g_{\phi\rm DR}\varphi_{\rm osc}\frac{2k}{m_\phi a}\left(\frac{a_{\rm osc}}{a}\right)^{\frac{3}{2}}\cos(2z)\right]\mu_{k\lambda}=0.
}
Comparing Eqs.~(\ref{generalmathieu}) and (\ref{mymathieu}) we find
\eq{
p=\frac{4k^2}{m_\phi^2 a^2},\quad q=\lambda g_{\phi\rm DR}\varphi_{\rm osc}\frac{2k}{m_\phi a}\left(\frac{a_{\rm osc}}{a}\right)^{\frac{3}{2}}.
}
The Mathieu equation is known to possess instability bands for certain values of $p$ and $q$ where the solution has an exponential growth. For $q\gg 1$ a large region of the parameter space is unstable, leading to broad parametric resonance. Nevertheless in this work we will focus in the narrow resonance regime, meaning  $q\ll 1$. In this case, parametric amplification occurs in the first resonance band given by $1-\lvert q\rvert <p < 1+\lvert q\rvert$
\eq{
1- g_{\phi\rm DR}\varphi_{\rm osc}\frac{2k}{m_\phi a}\left(\frac{a_{\rm osc}}{a}\right)^{\frac{3}{2}}<\frac{4k^2}{m_\phi^2 a^2}<1+g_{\phi\rm DR}\varphi_{\rm osc}\frac{2k}{m_\phi a}\left(\frac{a_{\rm osc}}{a}\right)^{\frac{3}{2}}.
}
Taking square root and using that $q$ is small
\eq{
1- g_{\phi\rm DR}\varphi_{\rm osc}\frac{k}{m_\phi a}\left(\frac{a_{\rm osc}}{a}\right)^{\frac{3}{2}}<\frac{2k}{m_\phi a}<1+ g_{\phi\rm DR}\varphi_{\rm osc}\frac{k}{m_\phi a}\left(\frac{a_{\rm osc}}{a}\right)^{\frac{3}{2}}.
}
In view of the previous result and using that $q$ is small, one can substitute $k/(m_\phi a)=1/2$ and then
\eq{
1-\frac{1}{2}g_{\phi\rm DR}\varphi_{\rm osc}\left(\frac{a_{\rm osc}}{a}\right)^{\frac{3}{2}}<\frac{2k}{m_\phi a}<1+\frac{1}{2}g_{\phi\rm DR}\varphi_{\rm osc}\left(\frac{a_{\rm osc}}{a}\right)^{\frac{3}{2}}. 
}
Hence, the condition for small $q$ or, equivalently, narrow resonance is
\eq{\label{narrowresonance}
g_{\phi\rm DR}\varphi_{\rm osc}\lesssim 1.
}
Consequently, the center of the instability band is at $k/a=m_\phi/2$ and the width is proportional to $g_{\phi\rm DR}\varphi_{\rm osc}$. Equivalently we can also substitute $a=2k/m_\phi$ to get
\eq{\label{resonantband}
1-\frac{1}{2}g_{\phi\rm DR}\varphi_{\rm osc}\left(\frac{a_{\rm osc}m_\phi}{2k}\right)^{\frac{3}{2}}<\frac{2k}{m_\phi a}<1+\frac{1}{2}g_{\phi\rm DR}\varphi_{\rm osc}\left(\frac{a_{\rm osc}m_\phi}{2k}\right)^{\frac{3}{2}}
}

Now we can estimate the enhancement that the mode function experiences due to parametric resonance that will translate into an enhancement on particle production. This can be computed as 
\eq{
\mu_{k\lambda}\propto \exp{\left(\left|\frac{q}{2}\right| z_{\rm res}\right)},\quad \textrm{where}\quad   z_{\rm res}=\frac{1}{2}m_\phi(t_{\rm end}-t_{\rm ini}),
}
and $t_{\rm init}$, $t_{\rm end}$ are the times in which a mode $k$ enters and leaves the resonant band, respectively. From Eq.~(\ref{resonantband}) we can obtain the size of the scale factor at those times
\eq{
a_{\rm ini}=\frac{2k}{m_\phi}\left[1-\frac{g_{\phi\rm DR}}{2}\varphi_{\rm osc}\left(\frac{a_{\rm osc}m_\phi}{2k}\right)^{\frac{3}{2}}\right],\quad a_{\rm end}=\frac{2k}{m_\phi}\left[1+\frac{g_{\phi\rm DR}}{2}\varphi_{\rm osc}\left(\frac{a_{\rm osc}m_\phi}{2k}\right)^{\frac{3}{2}}\right].
}
The ALP starts oscillating during radiation domination and continues oscillating during matter domination. For the modes that enter and leave the resonance band during radiation domination we get using Eq.~(\ref{tinraddom})
\begin{eqnarray}
  t_{\rm ini}&=&\frac{1}{2H_{\rm osc}}\left(\frac{2k}{a_{\rm osc}m_\phi}\right)^{2}\left[1-g_{\phi\rm DR}\varphi_{\rm osc}\left(\frac{a_{\rm osc}m_\phi}{2k}\right)^{\frac{3}{2}}\right],\nonumber \\
  t_{\rm end}&=&\frac{1}{2H_{\rm osc}}\left(\frac{2k}{a_{\rm DR}m_\phi}\right)^{2}\left[1+g_{\phi\rm DR}\varphi_{\rm osc}\left(\frac{a_{\rm osc}m_\phi}{2k}\right)^{\frac{3}{2}}\right].  
\end{eqnarray}
while for those modes that enter and leave the band during matter domination after using Eq.~(\ref{tinmatdom})
\begin{align}
  t_{\rm ini}&=&\frac{2}{3H_{\rm osc}}\left(\frac{a_{\rm eq}}{a_{\rm osc}}\right)^{\frac{1}{2}}\left(\frac{2k}{a_{\rm osc}m_\phi}\right)^{\frac{3}{2}}\left[1-\frac{3}{4}g_{\phi\rm DR}\varphi_{\rm osc}\left(\frac{a_{\rm osc}m_\phi}{2k}\right)^{\frac{3}{2}}\right]+t_1-\frac{2}{3H_{\rm osc}}\left(\frac{a_{\rm eq}}{a_{1}}\right)^{\frac{1}{2}},\nonumber \\
  t_{\rm end}&=&\frac{2}{3H_{\rm osc}}\left(\frac{a_{\rm eq}}{a_{\rm osc}}\right)^{\frac{1}{2}}\left(\frac{2k}{a_{\rm osc}m_\phi}\right)^{\frac{3}{2}}\left[1+\frac{3}{4}g_{\phi\rm DR}\varphi_{\rm osc}\left(\frac{a_{\rm osc}m_\phi}{2k}\right)^{\frac{3}{2}}\right]+t_1-\frac{2}{3H_{\rm osc}}\left(\frac{a_{\rm eq}}{a_{1}}\right)^{\frac{1}{2}}, 
\end{align}
where $a_{\rm eq}=\sqrt{\Omega^0_{\rm rad}/\Omega^0_{\rm mat}}$ is the value of the scale factor at matter-radiation equality.

Hence, $z_{\rm res}$ in radiation domination epoch is given by
\eq{
z_{\rm res}=\frac{m_{\phi}}{2H_{\rm osc}}g_{\phi\rm DR}\varphi_{\rm osc}\left(\frac{2k}{a_{\rm osc}m_\phi}\right)^{\frac{1}{2}}
}
while for modes entering in the resonance band during matter domination we get

\eq{
z_{\rm res}=\frac{m_{\phi}}{2H_{\rm osc}}\left(\frac{a_{\rm eq}}{a_{\rm osc}}\right)^{\frac{1}{2}}g_{\phi\rm DR}\varphi_{\rm osc}.
}
The total growth exponent for radiation domination modes is given by
\eq{
\left|\frac{ q}{2}\right| z_{\rm res}=\left(g_{\phi\rm DR}\varphi_{\rm osc}\right)^2\frac{m_\phi}{4H_{\rm osc}}\frac{a_{\rm osc}m_\phi}{2k},
}
while for matter domination modes
\begin{align}
  \left|\frac{q}{2}\right| z_{\rm res}=\left(g_{\phi\rm DR}\varphi_{\rm osc}\right)^2\frac{m_\phi}{4H_{\rm osc}}\left(\frac{a_{\rm eq}}{a_{\rm osc}}\right)^{\frac{1}{2}}\left(\frac{a_{\rm osc}m_\phi}{2k}\right)^{\frac{3}{2}}. 
\end{align}
Since for kinematics $m_\phi=2k/a\lesssim 2k/a_{\rm osc}$, the most enhanced mode is $k=a_{\rm osc} m_\phi /2$. The enhancement is proportional to the coupling of the ALP to photons, the displacement of the field with respect to to the minimum when the ALP starts oscillating, and the ratio between the ALP mass and the Hubble expansion rate at the time when oscillations begin. The modes that enter the resonance band during matter dominated epoch are less enhanced, due to the suppression term $a_{\rm eq}/a_{\rm osc}$. Notice that, during radiation domination epoch, the later the oscillations begin the more enhanced is the mode function for a fixed mass since the enhancement is proportional to $m_\phi/H_{\rm osc}$ and $H$ is decreasing in time ($\dot{H}<0$) in a radiation and matter dominated Universe. As a result, more dark photons are generated, leading to a reduced amount of dark matter today due to energy conservation. In principle, this allows for a higher $H_0$ and a lower $S_8$ compared to the values predicted by $\Lambda\rm CDM$, potentially addressing both tensions simultaneously.

\subsubsection{Analytical calculation}
Here we will obtain analytical expressions for the single particle occupation number and the comoving energy density of dark photons using the WKB approach. Dark Radiation is produced once the ALP starts its oscillatory phase, from $a=a_{\rm osc}$ onward. We will further neglect anharmonicity corrections.  

The starting point is Eq.~(\ref{betaeq}) when the ALP starts oscillating at $a=a_{\rm osc}$. Let us first compute the accumulated phase
\eq{\label{accumulatedphase}
\Psi_{k\lambda}=\int_{\eta_{\rm osc}}^{\eta}d\eta'\omega_{k\lambda}(\eta')=\int_{\eta_{\rm osc}}^{\eta}d\eta'\sqrt{k^2-k\lambda g_{\phi\rm DR}\varphi'}\approx \int_{\eta_{\rm osc}}^{\eta}d\eta'k=k\left(\eta-\eta_{\rm osc}\right),
}
where we have ignored at leading order the contribution proportional to $g_{\phi\rm DR}\varphi_{osc}$ coming from $\varphi'$ since we work in the narrow resonance regime given by Eq.~(\ref{narrowresonance}) and $k/a\approx m_\phi/2$ in the resonant band according to Eq.~(\ref{resonantband}), meaning that $q<1$.
On the other hand, we also need to compute the quantity
\eq{
\frac{\omega'_{k\lambda}}{2\omega_{k\lambda}}=\frac{-k\lambda g_{\phi\rm DR}\varphi''}{4\omega^2_{k\lambda}}=\frac{-k\lambda g_{\phi\rm DR}}{4(k^2-k\lambda g_{\phi\rm DR}\varphi')}\left(-2H a \varphi'-m_\phi^2a^2\varphi\right)\approx \frac{k\lambda g_{\phi\rm DR}m_\phi^2a^2\varphi}{4(k^2-k\lambda g_{\phi\rm DR}\varphi')},
}
where in the second equality we use the ALP eom (\ref{alpeom}) and in the last equality we neglected $H/m_\phi$ during oscillations. Using the explicit forms of $\varphi$ and $\varphi'$ from Eqs.~(\ref{alpposition}) and (\ref{alpvelocity}) neglecting $H/m_\phi$, we get
\eq{
\frac{\omega'_{k\lambda}}{2\omega_{k\lambda}}=\frac{k\lambda g_{\phi\rm DR} m_\phi^2a^2}{4\left[k^2-k\lambda g_{\phi\rm DR}\varphi_{\rm osc}a_{\rm osc}^{\frac{3}{2}}a^{-\frac{1}{2}}m_\phi\cos(m_\phi t(\eta)+\nu)\right]}\varphi_{\rm osc}\left(\frac{a_{\rm osc}}{a}\right)^{\frac{3}{2}}\sin(m_\phi t(\eta)+\nu).
}
Again, since in the resonant band $k/a\approx m_\phi/2$ and in the narrow resonance regime we have $g_{\phi\rm DR}\varphi_{\rm osc}<1$, we can neglect the second term of the denominator and obtain
\eq{
\frac{\omega'_{k\lambda}}{2\omega_{k\lambda}}=\frac{\lambda g_{\phi\rm DR} m_\phi^2a^2}{4k}\varphi_{\rm osc}\left(\frac{a_{\rm osc}}{a}\right)^{\frac{3}{2}}\sin(m_\phi t(\eta)+\nu).
}
Substituting this quantity and the accumulated phase (\ref{accumulatedphase}) into Eq.~(\ref{betaeq}) we obtain
\eq{
\beta_{k\lambda}=\frac{\lambda g_{\phi\rm DR} m_\phi^2}{8ik}\varphi_{\rm osc}a_{\rm osc}^{\frac{3}{2}}\int_{\eta_{\rm osc}}^{\eta}d\eta'a(\eta')^{\frac{1}{2}}\left(e^{i\psi_{k\lambda}^{-}(\eta')}-e^{-i\psi_{k\lambda}^{+}(\eta')}\right),
}
where we have written $\sin(m_\phi t +\nu)=\frac{e^{i(m_\phi t +\nu)}-e^{-i(m_\phi t +\nu)}}{2i}$ and the phases are defined as
\eq{
\psi_{k\lambda}^{\pm}(\eta')=\pm 2k(\eta'-\eta_{\rm osc})+m_\phi t(\eta')+\nu.
}
The integrand oscillates rapidly so we can apply the stationary phase principle. This establishes that an oscillatory integral of the form
\eq{
\int_{a}^b dx~g(x)e^{i f(x)}
}
where $f(x)$ and $g(x)$ are not oscillatory, is dominated by the points $x_0$ where the phase is stationary
\eq{
\left.\frac{d f}{dx}\right|_{x_0}=0, \quad x_0\in(a,b), \quad \left.\frac{d^2 f}{dx^2}\right|_{x_0}\neq 0
}
and the result of the integral is given, at leading order by
\eq{
\int_{a}^b dx~g(x)e^{i f(x)}= \sum_{x_0}g(x_0)e^{i f(x_0)+\textrm{sign}(f''(x_0))i\frac{\pi}{4}}\left(\frac{2\pi}{\lvert f''(x_0)\rvert}\right)^{\frac{1}{2}}.
}
Applying the stationary phase principle to our case, the integral is dominated by the instants when
\eq{
\left.\frac{d\psi^{-}_{k\lambda}}{d\eta}\right|_{\eta_k}=-2k+m_\phi a(\eta_k)=0\rightarrow a_k\equiv a(\eta_k)=\frac{2k}{m_\phi},\quad \left.\frac{d^2\psi^{-}_{k\lambda}}{d\eta^2}\right|_{\eta_k}=m_\phi a'_k=m_\phi \mathcal H(a_k)a_k,
}
and the result is given by
\eq{
\beta_{k\lambda}=\frac{\lambda g_{\phi\rm DR} m_\phi^2}{8ik}\varphi_{\rm osc}a_{\rm osc}^{\frac{3}{2}}a_k^{\frac{1}{2}}e^{i k(\eta_k-\eta_{\rm osc})+i\frac{\pi}{4}}\left(\frac{2\pi}{m_\phi \mathcal H(a_k)a_k}\right)^{\frac{1}{2}}.
}
Thus, the single particle occupation number reads 
\eq{\label{occupationnumberfinal}
n_{k\lambda}=\lvert\beta_{k\lambda}\rvert^2=\pi\frac{g^2_{\phi\rm DR}\varphi_{\rm osc}^2}{32k^2}\frac{a^3_{\rm osc}m_\phi^3}{\mathcal H(a_k)}\Theta\left(\frac{2k}{m_\phi}-a_{\rm osc}\right)\Theta\left(a-\frac{2k}{m_\phi}\right),
}
where $\Theta(x)$ is the Heaviside step function. 
These formulas take into account the stationary phase approximation because, at a fixed time $\eta$, the mode that is being created satisfies $k=1/2 m_\phi a(\eta)$. The first step function restricts every mode to be larger than $k>1/2 m_\phi a_{\rm osc}$ since the ALP starts oscillating at $a=a_{\rm osc}$. The second step function sets to zero the occupation number of a mode that at a time $\eta$ has not been already created.

We need to relate this expression to the energy density of the DR. On the one hand, the mixed energy momentum tensor for dark radiation is
\begin{align}
    (T_{\rm DR})^{\mu}_\nu=\rm diag\left(\rho,-\frac{1}{3}\rho,-\frac{1}{3}\rho,-\frac{1}{3}\rho\right),
\end{align}
where we used $\omega_{\rm rad}=1/3$. On the other hand, it can be written in terms of the single particle occupation number as 
\begin{align}\label{energymomentumtensorDR}
    T_{\rm DR}^{\mu\nu}=\sum_{\lambda}\int\frac{d^3P}{(2\pi)^3 P^0}\lvert g\rvert ^{-\frac{1}{2}}P^{\mu}P^{\nu}n_{k\lambda},
\end{align}
 where $P_{\mu}$ is the conjugated momentum of the coordinate $x^{\mu}$. Following Ref.~\cite{Ma:1995ey} 
we get the relation between the conjugate momentum and the momentum at a fixed spatial coordinate $p_i=p^i$. In general, one defines the action 
\begin{align}
    \mathcal L_{\rm DR} =\frac{1}{2}g_{\mu \nu} u_{\rm DR}^\mu u_{\rm DR}^\nu,
\end{align}
where $u_{\rm DR}^\mu =dx_{\rm DR}^\mu/d\tau $ is the dark photon 4-velocity. 
Then, the conjugated momentum is given by
\begin{align}
    P_\mu=\frac{\partial\mathcal L}{\partial u_{\rm DR}^\mu}=g_{\mu\nu}u_{\rm DR}^\nu\rightarrow P^\mu= u_{\rm DR}^\mu
\end{align}
For the zero-th component we have
\begin{align}
P^0=u^0_{\rm DR}.  
\end{align}
From the action above, the geodesic equation is thus
\begin{align}
    \frac{d u_{\rm DR}^\mu}{d\tau}+\Gamma^\mu_{\nu\rho}u_{\rm DR}^\nu u_{\rm DR}^\rho=0.
\end{align}
Using that 
\begin{align}
    \frac{d u^\mu_{\rm DR}}{d\tau}=\frac{d P^\mu}{d\tau}=P'^\mu u^0 =P^0 P'^\mu,
\end{align}
the geodesic equation becomes
\begin{align}\label{geodesicequation}
    P^0 P'^\mu + \Gamma^\mu_{\nu\rho}P^\nu P^\rho=0.
\end{align}
This together with the restriction of null-geodesic
\begin{align}
    g_{\mu\nu}u_{\rm DR}^\mu u_{\rm DR}^\nu=0\rightarrow g_{\mu\nu }P^\mu P^\nu =0 ,
\end{align}
determines the trajectories of dark photons in space-time. Using the FLRW metric, the condition of null-geodesic is given by
\begin{align}
    a^2 (P^0)^2-a^2\delta_{ij}P^i P^j=0.
\end{align}
At a fixed spatial coordinate the above expression becomes
\begin{align}
 (p^0)^2-\delta_{ij}p^ip^j=0,   
\end{align}
hence, identifying we get
\begin{align}
    &p^0=a P^0\\
    &p^i=a P^i
\end{align}
and $(p^0)^2=\sum_i(p^i)^2=\lvert\vec p\rvert^2$. To obtain the evolution of the quantity $p^0$ we need the Christoffel symbols of the FLRW metric in conformal time (see  Eq.~(\ref{christoffelbackgroundconformal})). Hence
\begin{align}
    \frac{p^0}{a} \left(\frac{p^0}{a}\right)' +\Gamma^0_{00} \left(\frac{p^0}{a}\right)^2+\Gamma^0_{ij}\frac{p^i }{a} \frac{p^j}{a}=0\rightarrow (a p^0)'=0, 
\end{align}
and then $\lvert \vec p\rvert\propto a^{-1}$. For that purpose, we define $k^0 \equiv k= a p^0$, $k^i=a p^i$ with $k^2=\sum_i (k^i)^2 $ so that $k$ is a constant. Then Eq.~(\ref{energymomentumtensorDR}) becomes
\begin{align}
    T_{\rm DR}^{\mu\nu}=\sum_\lambda\int\frac{d^3k}{(2\pi)^3 \frac{k}{a^2}}a^{-4}\frac{k^\mu}{a^2} \frac{k^\nu}{a^2}n_{k\lambda}=a^{-6}\sum_\lambda\int\frac{d^3k}{(2\pi)^3}\frac{k^\mu k^\nu}{k} n_{k\lambda}.
\end{align}
For the $T^{00}_{\rm DR}$ component we have
\begin{align}
T_{\rm DR}^{00}=a^{-6} \sum_\lambda\int  \frac{d^3k}{(2\pi)^3} k~n_{k\lambda}.
\end{align}
The energy density is given by the time-time component of the mixed tensor. Therefore,
\begin{align}
   \rho_{\rm DR}=(T_{\rm DR})^0_0=g_{00}T^{00}_{\rm DR}=a^2a^{-6} \sum_\lambda\int  \frac{d^3k}{(2\pi)^3} k~n_{k\lambda}=a^{-4}\sum_\lambda\int  \frac{d^3k}{(2\pi)^3} k ~n_{k\lambda}.
\end{align}
Computing the integral
\begin{eqnarray}
    a^4(\eta)\rho_{\rm DR}(\eta) &=& 2\int_0^\infty \frac{4\pi}{(2\pi)^3}dk~k^2~k~ \frac{\pi}{32k^2}\left(g_{\phi\rm DR}\varphi_{\rm osc}\right)^2 \frac{m^3_\phi a^3_{\rm osc}}{\mathcal H(a_k)}\Theta\left(\frac{2k}{m_\phi}-a_{\rm osc}\right)\Theta\left(a-\frac{2k}{m_\phi}\right)\nonumber \\
    &=& \frac{\left(g_{\phi\rm DR}\varphi_{\rm osc}\right)^2}{32\pi}m^3_\phi a^3_{\rm osc}\int_{\frac{1}{2}m_\phi a_{\rm osc}}^{\frac{1}{2}m_\phi a(\eta)} dk~\frac{k}{\mathcal H(a_k)}.
\end{eqnarray}
Taking the derivative with respect to conformal time, applying the Fundamental Theorem of Calculus, and multiplying by $a^{-4}$, we obtain
\begin{align}
    a^{-4}\frac{d}{d\eta}(a^4\rho_{\rm DR})=\frac{\left(g_{\phi\rm DR}\varphi_{\rm osc}\right)^2}{32\pi}m^3_\phi a^3_{\rm osc}\frac{1}{2}m_{\phi}a\frac{1}{2}m_{\phi} a\mathcal H \frac{a^{-4}}{\mathcal H}=\frac{\left(g_{\phi\rm DR}\varphi_{\rm osc}\right)^2}{128\pi}m^4_\phi a\left(\frac{a_{\rm osc}}{a}\right)^3 .
\end{align}
Using Eq.~(\ref{ALPenergydensity}) for the energy density of the ALP during oscillations, this expression becomes
\begin{align} \label{DRevolution1}
 a^{-4}\frac{d}{d\eta}(a^4\rho_{\rm DR})=\frac{g^2_{\phi\rm DR}m^2_{\phi}}{64\pi}a\rho_{\phi}.
\end{align}
From this expression, we can obtain the decay width of the ALP. To do that, we parametrize the energy transfer between the ALP and the DR as
\begin{align}
    \nabla_\mu T_{\phi}^{\mu\nu}&=-Q^{\nu}\label{nablamuphi}\\
    \nabla_\mu T^{\mu\nu}_{\rm DR}&=Q^{\nu}\label{nablamurad},
\end{align}
where $Q^{\nu}$ is defined as in Refs.~\cite{Majerotto:2009np,Clemson:2011an,Valiviita:2015dfa}
\begin{align}
    Q^{\mu}=\Gamma_\phi(\rho_{\phi}+P_\phi) u^{\mu}_{\phi},
\end{align}
and $u^{\mu}_{\phi}$ is the ALP four-velocity. Notice that due to the interactions, none of the above energy momentum tensors is individually conserved, only its sum. In the ALP comoving frame only the zero component contributes $u^0_{\phi}=1/a$ so that $g_{\mu\nu}u_{\phi}^\mu u^\nu_{\phi}=1$. Besides, we shown that $P_\phi=0$ during oscillations. Using the Christoffel symbols in Eq.~(\ref{christoffelbackgroundconformal}), the left-hand side of Eq.~(\ref{nablamurad}) is
\begin{align}
    \nabla_\mu (T_{\rm DR})^{\mu}_0=\rho'_{\rm DR}+4\mathcal H\rho_{\rm DR}.
\end{align}
Hence
\begin{align}\label{DRevolution2}
  \rho'_{\rm DR}+4\mathcal H\rho_{\rm DR}=a\Gamma_\phi\rho_\phi\rightarrow a^{-4}\frac{d}{d\eta}(a^4\rho_{\rm DR})=a\Gamma\rho_\phi.  
\end{align}
Comparing the expressions in Eqs.~(\ref{DRevolution1}) and (\ref{DRevolution2}), we obtain the ALP decay width
\begin{align}\label{ALPdecaywidth}
\Gamma_\phi=\frac{g^2_{\phi\rm DR}m^3_\phi}{64\pi},   
\end{align}
result that matches the one computed using QFT. Finally, if we use the change of variables $d\eta=da/(a\mathcal H)$, we get
\begin{align}
    \frac{d}{da}(a^4\rho_{\rm DR})=a^4\frac{\Gamma_\phi}{\mathcal H}\rho_\phi=a^3\frac{\Gamma_\phi}{H}\rho_\phi.
\end{align}
This means that the production of dark radiation is only effective from the moment in which $\Gamma_\phi/H>1$. Since we only want to change the physics from recombination onward, $\Gamma_\phi/H_{\rm rec}\leq 1$, placing an upper bound to $\Gamma_\phi$ and, consequently, to the DR coupling $g_{\phi\rm DR}$. 

Analogously, for the ALP that behaves as CDM for $a>a_{\rm osc}$, as previously shown, we have the averaged energy-momentum tensor
\begin{align}
    (T_\phi)^{\mu}_\nu=\rm diag(\rho_\phi,0,0,0).
\end{align}
Its corresponding evolution is thus given by
\begin{align}
    \rho'_\phi+3\mathcal H\rho_\phi=-a\Gamma_\phi \rho_\phi\rightarrow a^{-3}\frac{d}{d\eta}(a^3\rho_\phi)=-a\Gamma_\phi\rho_\phi.
\end{align}
Hence, at the background level for $a>a_{\rm osc}$, this model evolves as the decaying cold DM to dark radiation scenario \cite{McCarthy:2022gok}. The main differences arise at the level of perturbations.

\subsection{Perturbative cosmology}\label{subsec:pert_cosmology}
The perturbations are evolved using the code CLASS \cite{Diego_Blas_2011}. We will follow Ref.~\cite{Ma:1995ey} to present the main equations.

In the synchronous gauge and using conformal time, the FLRW metric takes the form
\begin{align}
    ds^2=a(\eta)^2\left[d\eta^2-\left(\delta_{ij}+h_{ij}(x)\right)dx^i dx^j\right].
\end{align}
The non-vanishing Christoffel symbols are thus given by
\begin{align}\label{perturbedchristoffels}
    &\Gamma^{\eta}_{\eta\eta}=\mathcal{H};\\
    &\Gamma^{\eta}_{ij}=\mathcal{H}(\delta_{ij}+h_{ij})+\frac{1}{2} h'_{ij}\\
    &\Gamma^{i}_{j\eta}=\mathcal{H}\delta^i_j+\frac{1}{2} h'^i_j\\
    &\Gamma^i_{jk}=\frac{1}{2}\delta^{il}\left(\partial_j h_{lk}+\partial_k h_{jl}-\partial_l h_{jk}\right),
\end{align}
where the prime denotes derivatives with respect to conformal time and $\mathcal{H}=H a$. In Fourier space $h_{ij}(x)$ can be decomposed as
\begin{align}
h_{ij}(\eta,\vec{x})=\int\frac{d^3 q}{(2\pi)^3}e^{i \vec{q}\cdot \vec{x}}\left[\frac{q_i q_j}{q^2}h(\eta,\vec{q})+\left(\frac{q_i q_j}{q^2}-\frac{1}{3}\delta_{ij}\right)6\upsilon(\eta,\vec{q})\right],    
\end{align}
such that $h$ is used to denote the trace of $h_{ij}$ in both the real space and the Fourier space and $\upsilon$ is the traceless part of the metric perturbation. Since $\upsilon$ has no direct counterpart defined in real space, we do not place a tilde over it.

\subsubsection{ALP perturbations}

From now on, background quantities will be denoted by an overbar. From the Einstein equations, the ALP energy-momentum tensor is obtained from the action by
\begin{align}\label{ALPtensor}
    (T_\phi)_{\mu\nu}=-\frac{2}{\sqrt{\lvert g\rvert}}\frac{\partial(\sqrt{\lvert g\rvert}\mathcal{L}_\phi)}{\partial g^{\mu\nu}},
\end{align}
where the ALP Lagrangian is given by
\begin{align}
    \mathcal{L}_\phi= \frac{1}{2}\partial_\mu\phi\partial^\mu\phi-V(\phi)
\end{align}
and, at linear level in perturbations, $\phi=\overline\phi+\delta\phi$. 

We model the ALP as a fluid. The corresponding energy-momentum tensor is therefore written as
\begin{align}\label{fluidtensor}
   (T_\phi)_{\mu\nu}=(\rho_\phi+P_\phi)(u_\phi)_\mu (u_\phi)_\nu-P_\phi g_{\mu\nu}
\end{align}
where the four-velocity $u_\phi^\mu$ in these coordinates is
\begin{align}
    u^\mu_\phi= \left(\frac{1}{a},\frac{v^i_\phi}{a}\right),
\end{align}
and $v^i_\phi=(v_\phi)_i$ is the peculiar ALP fluid velocity measured by a local inertial observer with line element $ds^2=dt^2-\delta_{ij}dx^idx^j$. Linearizing Eq.~(\ref{fluidtensor}), the mixed tensor takes the form
\begin{align}
    (T_\phi)^{\mu}_\nu=\begin{pmatrix}
      \overline{\rho}_\phi+\delta\rho_\phi & -(\overline{\rho}_\phi+\overline{P}_\phi) (v_\phi)_i\\
      \left(\overline{\rho}_\phi+\overline{P}_\phi\right)v_\phi^i &-(\overline{P}_\phi+\delta P_\phi)\delta^i_j
    \end{pmatrix}.
\end{align}
Comparing with the linearized version of Eq.~(\ref{ALPtensor}), one gets a relation between the ALP field and the fluid quantities
\begin{align}\label{background_density_pressure}
       \overline{\rho}_\phi=\frac{1}{2a^2}\overline{\phi}'^2+V(\overline{\phi}),\quad \overline{P}_\phi=\frac{1}{2a^2}\overline{\phi}'^2-V(\overline{\phi})    \end{align}
for the background quantities and
\begin{align}\label{perturbed_density_pressure}
 \delta\rho_\phi=\frac{1}{a^2}\overline{\phi}'\delta\phi'+V'(\overline{\phi})\delta\phi,\quad \delta P_\phi=  \frac{1}{a^2}\overline{\phi}'\delta\phi'-V'(\overline{\phi})\delta\phi, \quad (v_\phi)_i= -\frac{\partial_i\delta\phi}{\overline{\phi}'}
\end{align}
for the linear perturbations. \footnote{A scalar field has vanishing shear $\Pi^{ij}_\phi$ at the linear level.} By linearizing the conservation equation in Eq.~(\ref{nablamuphi}) and employing the energy-momentum tensor defined in Eq.~(\ref{fluidtensor}), we obtain the evolution equation for the background ALP energy density
\begin{align}\label{background_density_eom}
    \overline{\rho}'_\phi+(3\mathcal{H}+a\Gamma^{\rm eff}_\phi)(\overline{\rho}_\phi+\overline{P}_\phi)=0, \quad\textrm{with } \Gamma^{\rm eff}_\phi(a)=\Gamma_\phi \Theta (a-a_{\rm osc}).
\end{align}
Using the relations in Eq.~(\ref{background_density_pressure}), we derive the background ALP eom in cosmic time, now including the effect of the decay width
\begin{align}\label{backgroundalpeom}
\ddot{\overline \phi}+(3H+\Gamma^{\rm eff}_\phi)\dot{\overline\phi}+V'(\overline\phi)=0.   
\end{align}
For the linear perturbations, we work in Fourier space. To avoid confusion between quantities in real space, ($m(x)$), and their Fourier counterparts, ($\widetilde m(q)$), we adopt the notation
\begin{align}
    m(\vec x)=\int\frac{d^3q}{(2\pi)^3}e^{i\vec q\cdot \vec x}\widetilde{m}(\vec q),
\end{align}
unless otherwise stated. 
Hence,
\begin{align}
&\tilde{\delta}'_\phi+\left(1+\frac{\overline{P}_\phi}{\overline\rho_\phi}\right)\left(\tilde{\theta}_\phi+\frac{1}{2}h'\right)+\left(3\mathcal{H}+a\Gamma^{\rm eff}_\phi\right)\left(\frac{\widetilde{\delta P}_\phi}{\widetilde{\delta\rho}_\phi}-\frac{\overline P_\phi}{\overline\rho_\phi}\right)\widetilde{\delta}_\phi=0\\
&\widetilde{\theta}_\phi '+\left(\mathcal{H}+\frac{\overline{P}_\phi'}{\overline{\rho}_\phi+\overline{P}_\phi}\right)\widetilde{\theta}_\phi-\frac{q^2\widetilde{\delta P}_\phi}{\overline\rho_\phi+\overline P_\phi}=0,
\end{align}
where $\widetilde\theta_\phi$ is the Fourier transform of the velocity divergence in real space $\theta_\phi=\partial_i v^i_\phi$ that translates into $\widetilde{\theta}_\phi=i q_i \widetilde{v}^i_\phi$ in Fourier space, and $\widetilde{\delta}_\phi=\widetilde{\delta\rho}_\phi/\overline\rho_\phi$ is the density contrast. These equations are valid in all regimes. Using the expressions for the perturbed density and pressure in Eq.~(\ref{perturbed_density_pressure}), we get the equation of motion for the perturbed ALP field. In cosmic time

\begin{align}\label{perturbed_ALP_eom}
\ddot{\widetilde{\delta\phi}}
+ (3H+\Gamma^{\rm eff}_\phi)\dot{\widetilde{\delta\phi}}
+\left(V''(\overline\phi)+\frac{q^2}{a^2}\right)\widetilde{\delta\phi}
+\frac{1}{2}\dot h\dot{\overline{\phi}}
=0,
\end{align}
that is the equation of a damped oscillator with a time dependent frequency and a source term proportional to $\dot h\dot{\overline\phi}$.

Now we need an equation that relates the density perturbations to the energy and velocity perturbations. Following Ref.~\cite{Unnikrishnan:2024agf}, we propose this relation in Fourier space
\begin{align}\label{deltap}
    \widetilde{\delta P}_\phi=c_1\widetilde{\delta\rho}_\phi-c_2(\overline{\rho}_\phi+\overline{p}_\phi)\frac{\widetilde{\theta}_\phi}{q^2}
\end{align}
where $c_1=c_e^2$ with $c_e$ the effective sound speed of the fluid measured in its rest frame, $c_1=\widetilde{\delta P}_\phi/\widetilde{\delta\rho}_\phi\rvert_{ v^i_\phi=0}$.
Under an infinitesimal gauge transformation, $x^\mu\rightarrow x^\mu+\xi^\mu$ 
with $\xi^\mu=(T,L^i)$, 
the magnitudes in Fourier space transform as
\begin{align}
    &\widetilde{\delta\rho}_\phi\rightarrow \widetilde{\delta\rho}_\phi -\overline\rho_\phi ' \widetilde T\\
    &\widetilde{\delta P}_\phi\rightarrow \widetilde{\delta P}_\phi-\overline{P}_\phi'\widetilde{T}\\
    &\widetilde \theta_\phi \rightarrow \widetilde \theta_\phi-q^2 \widetilde T\\
    &\widetilde \Pi ^{ij}_\phi\rightarrow \widetilde\Pi^{ij}_\phi,
\end{align}
where we took into account that the magnitudes transform because of the energy-momentum transformation and also because their arguments also transform. Consequently, Eq.~(\ref{deltap}) is gauge invariant if $c_2$ and is related to $c_1$ as
\begin{align} \label{gauge_invariant_c2}
    c_2=(3\mathcal{H}+a\Gamma^{\rm eff}_\phi)\left(\frac{\overline{P}'_\phi}{\overline{\rho}'_\phi}-c_1\right),
\end{align}
where the eom~(\ref{background_density_eom}) was used.
In our case, the background density and pressure are related through $P_\phi=\omega_{\phi}\rho_\phi$ before the onset of the oscillations, when $a<a_{\rm osc}$ and $\Gamma^{\rm eff}_\phi(a)=0$. Hence, the above expression simplifies
\begin{align}
    c_2=3\mathcal{H}\left(\omega_{\phi}-c_1\right).
\end{align}
In particular, during the phase of kinetic misalignment induced by the change in the slope of the potential, $ a_{\rm tr}\leq a< a_{\rm osc}$, we have $\omega_{\phi}=1$ and
\begin{align}
    c_2=3\mathcal{H}\left(1-c_1\right).
\end{align}
We need an expression for $c_1$. This is easy to obtain using the expressions in Eq.~(\ref{perturbed_density_pressure}) in Fourier space. From the velocity, we get
\begin{align}\label{velocity_divergence}
    (\widetilde{v_\phi})_i=-iq_i\frac{\widetilde{\delta\phi}}{\overline\phi'}\rightarrow \widetilde{\delta\phi}=i\frac{q^i (\widetilde{v_\phi})_i}{q^2}\overline\phi'=\frac{\widetilde\theta_\phi}{q^2}\overline\phi'=a^2\frac{\overline\rho_\phi+\overline P_\phi}{\overline\phi'}\frac{\widetilde\theta_\phi}{q^2}
\end{align}
where in the third step we used the definition of $\widetilde\theta_\phi$ and in the last step we multiplied and divided by $\overline\phi'$ and rewrote $\overline\phi'^2$ in the numerator in terms of the background pressure and density using Eq.~(\ref{background_density_pressure}).  Now, from Eq.~(\ref{perturbed_density_pressure}), it is clear that the perturbed density and pressure satisfy
\begin{align}
    \widetilde{\delta P}_\phi=\widetilde{\delta\rho}_\phi-2V'(\overline\phi)\widetilde{\delta\phi}=\widetilde{\delta\rho}_\phi-2\frac{V'(\overline\phi)}{\overline\phi'}a^2(\overline\rho_\phi+\overline P_\phi)\frac{\widetilde\theta_\phi}{q^2}.
\end{align}
The quantity $2a^2V'(\overline\phi)/\overline\phi'$ can be obtained from the derivatives of the background pressure and density with respect to conformal time
\begin{align}
    2a^2\frac{V'(\overline\phi)}{\overline\phi'^2}\overline \phi'=\frac{\overline\rho_\phi'-\overline P_\phi'}{(\overline\rho_\phi+\overline P_\phi)}=3\mathcal H \left(\frac{\overline P_\phi'}{\overline\rho_\phi'}-1\right),
\end{align}
where again we multiplied the numerator and denominator by $\overline\phi'$ and used Eq.~(\ref{background_density_pressure}), and in the last step we factored out $\overline\rho_\phi'$ and used its eom (\ref{background_density_eom}). Hence
\begin{align}
  \widetilde{\delta P}_\phi= \widetilde{\delta\rho}_\phi-3\mathcal H \left(\frac{\overline P_\phi'}{\overline\rho_\phi'}-1\right)(\overline\rho_\phi+\overline P_\phi)\frac{\widetilde\theta_\phi}{q^2}.
\end{align}
Comparing with Eq.~(\ref{deltap}), we obtain 
\begin{align}
    c_1=1,\quad c_2=3\mathcal H \left(\frac{\overline P_\phi'}{\overline\rho_\phi'}-1\right),
\end{align}
in agreement with Ref.~\cite{Unnikrishnan:2024agf}. Clearly, during the phase with constant $\omega_\phi$ $c_2=3\mathcal H\left(w_\phi-1\right)$, while during the kinetic misalignment phase $c_2=0$.

Finally, during the oscillation phase $a\geq a_{\rm osc}$, the averaged background pressure is $\overline{P}_\phi=0$ and $\Gamma^{\rm eff}_\phi(a)=\Gamma_\phi$. During this stage, the ALP oscillates quickly around a minimum of its potential, provided that $m_\phi>3H$. To take care of these oscillations, we need to average to obtain a new expression relating $\widetilde {\delta P_\phi}$, $\widetilde{\delta \rho_\phi}$ and $\widetilde{\theta_\phi}$. This can be done as follows. The leading order solution for the background ALP in Eq.~(\ref{backgroundalpeom}),
expanded around a minimum of the cosine potential and in the limits of small
 $\dot H/m_\phi^2$, $H/m_\phi$, $\Gamma_\phi/m_\phi$, including leading-order amplitude corrections from $\Gamma_\phi$, is given by
\begin{align}
\varphi(t)=\varphi_0a^{-\frac{3}{2}}e^{-\Gamma_\phi t/2}\cos(m_\phi t),  
\end{align}
where we set the initial phase to zero without loss of generality. The derivative with respect to cosmic time is
\begin{align}
\dot{\varphi}(t)=-\varphi_0a^{-\frac{3}{2}}e^{-\Gamma_\phi t/2}\left(\frac{1}{2}\left(3H+\Gamma_\phi\right)\cos(m_\phi t)+m_\phi\sin(m_\phi t)\right).    
\end{align}
On the other hand, ignoring the source term in the perturbed ALP eom in Eq.~(\ref{perturbed_ALP_eom}) and setting $V''(\overline{\phi})=m_\phi^2$, we have
\begin{align}
    \widetilde{\delta\phi}(t)= A_q a^{-\frac{3}{2}}e^{-\Gamma_\phi t/2}\cos\left(\omega_q t\right), \quad \omega_q=\sqrt{m_\phi^2+\frac{q^2}{a^2}}
\end{align}
where $A_q$ is only a function of the mode $q$, we used the leading order of the WKB expansion of the solution and imposed that the background and perturbed ALP fields are aligned at the moment when oscillations begin. The derivative with respect to cosmic time is
\begin{align}
\dot{\widetilde{\delta{\phi}}}(t)=-A_qa^{-\frac{3}{2}}e^{-\Gamma_\phi t/2}\left(\frac{1}{2}\left(3H+\Gamma_\phi\right)\cos(\omega_q t)+\omega_q\sin(\omega_q t)\right),    
\end{align}
where we neglected $\dot{\omega}_q$ at leading order in $H/m_\phi$ since $\dot\omega_q=\mathcal{O}(H/m_\phi)$, and thus $\omega_q$ is slowly varying.

To compute the pressure, density, and velocity divergence, we need to average over the high frequencies. From Eqs.~(\ref{perturbed_density_pressure}) and (\ref{velocity_divergence}) in cosmic time, we need to calculate the average of the following quantities
\begin{align}
\dot{\varphi}\dot{\widetilde{\delta\phi}} &=
\varphi_0 A_q a^{-3} e^{-\Gamma_\phi t}
\Bigg[
m_\phi\omega_q\sin(\omega_q t)\sin(m_\phi t)+ \nonumber \\
&\quad + \frac{1}{2}\left(3H+\Gamma_\phi\right)
\Big(
\omega_q\sin(\omega_q t)\cos(m_\phi t)
+ m_\phi\cos(\omega_q t)\sin(m_\phi t)
\Big)
\Bigg]
\label{1steq}\\
\varphi\widetilde{\delta\phi} &=
\varphi_0 A_q a^{-3}e^{-\Gamma_\phi t}\cos(\omega_q t)\cos(m_\phi t)\\
\dot{\varphi}\widetilde{\delta\phi} &=
-\varphi_0 A_q a^{-3}e^{-\Gamma_\phi t}m_\phi\cos(\omega_q t)\sin(m_\phi t).
\end{align}
To illustrate how we take the average, let us focus, for instance, on the first term of equation Eq.~(\ref{1steq}). Using the relations for the product of two trigonometric functions, we have
\begin{align}
    \sin(\omega_q t)\sin(m_\phi t)=-\frac{1}{2}\left[\cos((\omega_q+m_\phi)t)-\cos((\omega_q-m_\phi)t)\right].
\end{align}
Clearly, the high frequency is $\omega_q+m_\phi$. Since $\omega_q$ is slowly varying, we can define a period $T=2\pi/(\omega_q+m_\phi)$. Hence, assuming that the scale factor, $a$, and $e^{-\Gamma_\phi t}$ are also slowly varying functions due to Eq.~(\ref{change_scale_factor}) and the assumption $\Gamma_\phi/m_\phi<1$, we have 
\begin{align}
    \left\langle  \sin(\omega_q t)\sin(m_\phi t) \right\rangle_T=&\frac{\omega_k+m_\phi}{2\pi}\left(-\frac{1}{2}\right)\int_0^{\frac{2\pi}{\omega_k+m_\phi}}\left[\cos((\omega_q+m_\phi)t)-\cos((\omega_q-m_\phi)t)\right]dt=\nonumber\\=&-\frac{1}{4\pi}\frac{\omega_k+m_\phi}{\omega_k-m_\phi}\sin\left(\frac{4\pi m_\phi}{\omega_k+m_\phi}\right).
\end{align}
Repeating the same process for all the trigonometric combinations above, we get
\begin{align}
\left\langle\dot{\varphi}\dot{\widetilde{\delta\phi}}\right\rangle_T
&= \varphi_0 A_q a^{-3} e^{-\Gamma_\phi t}\frac{1}{4\pi}(\omega_q+m_\phi)
\Bigg[
-\frac{m_\phi\omega_q}{\omega_q-m_\phi}
\sin\left(\frac{4\pi m_\phi}{\omega_q+m_\phi}\right)+
\nonumber \\
&\quad + \frac{1}{2}\left(3H+\Gamma_\phi\right)
\left(1-\cos\left(\frac{4\pi m_\phi}{\omega_q+m_\phi}\right)\right)
\Bigg] \\
\left\langle \varphi\widetilde{\delta\phi}\right\rangle_T
&= -\varphi_0 A_q a^{-3}e^{-\Gamma_\phi t}\frac{1}{4\pi}
\frac{\omega_q+m_\phi}{\omega_q-m_\phi}
\sin\left(\frac{4\pi m_\phi}{\omega_q+m_\phi}\right)\\
\left\langle \dot{\varphi}\widetilde{\delta\phi}\right\rangle_T
&= \varphi_0 A_q a^{-3}e^{-\Gamma_\phi t}m_\phi\frac{1}{4\pi}
\frac{\omega_q+m_\phi}{\omega_q-m_\phi}
\left[1-\cos\left(\frac{4\pi m_\phi}{\omega_q+m_\phi}\right)\right].
\end{align}
Substituting these expressions into Eqs.~(\ref{perturbed_density_pressure}) and (\ref{velocity_divergence}), the density, pressure and velocity divergence read
\begin{align}
\left\langle \widetilde{\delta\rho}_\phi\right\rangle_T
&= \varphi_0 A_q a^{-3} e^{-\Gamma_\phi t}\frac{1}{4\pi}
\Bigg[
- m_\phi\frac{(\omega_q+m_\phi)^2}{\omega_q-m_\phi}
\sin\left(\frac{4\pi m_\phi}{\omega_q+m_\phi}\right)+
\nonumber \\
&\quad + \frac{1}{2}\left(3H+\Gamma_\phi\right)(\omega_q+m_\phi)
\left(1-\cos\left(\frac{4\pi m_\phi}{\omega_q+m_\phi}\right)\right)
\Bigg] \\
\left\langle \widetilde{\delta P}_\phi\right\rangle_T
&= \varphi_0 A_q a^{-3} e^{-\Gamma_\phi t}\frac{1}{4\pi}(\omega_q+m_\phi)
\Bigg[
- m_\phi \sin\left(\frac{4\pi m_\phi}{\omega_q+m_\phi}\right)+
\nonumber \\
&\quad + \frac{1}{2}\left(3H+\Gamma_\phi\right)
\left(1-\cos\left(\frac{4\pi m_\phi}{\omega_q+m_\phi}\right)\right)
\Bigg] \\
\left\langle\left(\overline \rho_\phi+\overline P_\phi\right)\frac{\theta_\phi}{q^2} \right\rangle_T
&= \frac{1}{a}\varphi_0 A_q a^{-3}e^{-\Gamma_\phi t}m_\phi\frac{1}{4\pi}
\frac{\omega_q+m_\phi}{\omega_q-m_\phi}
\left[1-\cos\left(\frac{4\pi m_\phi}{\omega_q+m_\phi}\right)\right].
\end{align}
Combining these expressions, we obtain the relation
\begin{align}\label{relation_pressure_density_velocity_oscillations}
 \left\langle \widetilde{\delta P}_\phi\right\rangle_T=\frac{\omega_q-m_\phi}{\omega_q+m_\phi} \left\langle \widetilde{\delta\rho}_\phi\right\rangle_T+(3\mathcal H+a\Gamma_\phi)\frac{\omega_q-m_\phi}{\omega_q+m_\phi} \left\langle\left(\overline \rho_\phi+\overline P_\phi\right)\frac{\theta_\phi}{q^2} \right\rangle_T. 
\end{align}
This can be compared to Eq.~(\ref{deltap}), which allows us to identify the effective sound speed
\begin{align}\label{scalar_field_sound_speed}
    c_1=\frac{\omega_q-m_\phi}{\omega_q+m_\phi}=\frac{q^2}{4m_\phi^2 a^2}\left(\frac{1}{2}+\frac{q^2}{4m_\phi^2 a^2}+\sqrt{\frac{1}{4}+\frac{q^2}{4m_\phi^2 a^2}}\right)^{-1},
\end{align}
expression that is valid for all scales. In particular
\begin{align}
 c_1\approx \begin{cases}
     \frac{q^2}{4m_\phi^2a^2}, &\frac{q}{ m_\phi a}\rightarrow 0 \\
     1, &\frac{q}{m_\phi a}\rightarrow \infty,
 \end{cases}   
\end{align}
in agreement with Ref.~\cite{Hu:2000ke}. This differs from the standard DM scenario, where $c_1=0$, since DM is also a perfect fluid at the perturbation level. On the other hand, our full expression does not coincide with those of Refs.~\cite{Hwang:2009js,Passaglia:2022bcr}. However, according to Ref.~\cite{Hu:2000ke}, the details of the interpolation between these limits are not physically relevant. For $c_2$ we obtained
\begin{align}
    c_2=-(3\mathcal H+a\Gamma_\phi)\frac{\omega_q-m_\phi}{\omega_q+m_\phi},
\end{align}
which is related to $c_1$ as in Eq.~(\ref{gauge_invariant_c2}) with $\overline{P}_\phi=0$. This is a manifestation of the gauge invariance of Eq.~(\ref{relation_pressure_density_velocity_oscillations}).

\subsubsection{DR perturbations}

Dark radiation evolves according to the Boltzmann equation
\begin{align}
    \frac{d f(\eta,\vec x_1,\vec p_1)}{d\eta}=a C(\eta,\vec x_1,\vec p_1),
\end{align}  
where $f(\eta,\vec x_1,\vec p_1)$ is the distribution function of dark photons which depends on time, the position and the 3-momentum; and $C(\vec p_1)$ is the collision term. For a two body decay in the rest frame of a local observer at a fixed spatial coordinate, the collision term takes the form
\begin{align}
C(\vec p_1)=\frac{1}{2|\vec p_1|}\int d\Pi_\phi\, d\Pi_2\,|\mathcal M|^2\, g(\vec p_\phi)\,(1+f(\vec{p}_1))(1+f(\vec{p}_2))\,(2\pi)^4\delta^{(4)}(p_\phi-p_1-p_2),
\end{align}
with
\begin{align}
    d\Pi_i=\frac{d^3\vec{p}_i}{(2\pi)^32E_i(\vec{p}_i )}.
\end{align}
Here $p_\phi$ and $p_{1,2}$ denote the four-momenta of the ALP and DR, respectively, at a fixed spatial coordinate; $\lvert \mathcal{M}\rvert^2$ is squared matrix element of the process and $g(\vec q)$ is the ALP distribution function. We will consider that $g(\vec q)$ takes the form
\begin{align}
    g(\vec p_\phi)=\frac{\rho_\phi + P_\phi}{m_\phi} (2\pi)^3\delta^{(3)}(\vec p_\phi - m_\phi \vec v_\phi),
\end{align}
where $\vec v_\phi =\vec v_{\phi}(x)$. If there is no stimulated emission of DR, then $1+f(\vec p_i)\approx 1$ and the expression simplifies
\begin{align}
 C(\vec p_1)=\frac{1}{2\lvert \vec p_1\rvert}\int \frac{d^3 \vec p_\phi}{(2\pi)^3 2E_\phi(\vec{p}_\phi)}\frac{d^3\vec p_2}{(2\pi)^3 2 E_2(\vec{p}_2)}\lvert\mathcal M\rvert^2 g(\vec{p}_\phi)(2\pi)^4\delta^{(4)}(p_\phi-p_1-p_2).   
\end{align}
We would like to write the Boltzmann equation in terms of the new momentum variables $k_i$. Following the method we used above but with the perturbed metric, one first obtains the relation between the conjugated momentum $P^\mu$ and $p^0$, $p^i$. From the null-geodesic condition
\begin{align}
    a^2 (P^0)^2-a^2(\delta_{ij}+h_{ij}(x))P^i P^j=0
\end{align}
we get
\begin{align}
    &P^0=\frac{p^0}{a},\quad P^i = \left(\delta^i_j-\frac{1}{2}h^i_j(x) \right)\frac{p^j}{a}.  
\end{align}
As before we introduce $k^0=a p^0$ and $k^i=a p^i$, so that
\begin{align}
 &P^0=\frac{k^0}{a^2},\quad P^i = \left(\delta^i_j-\frac{1}{2}h^i_j(x) \right)\frac{k^j}{a^2}.    
\end{align}
and $(k^0)^2=(k)^2=\sum_i(k^i)^2$. In this case, due to the perturbation $h_{ij}$ in the metric, $k$ is not conserved but evolves according to the time component of the geodesic equation in Eq.~(\ref{geodesicequation})
\begin{align}\label{kprime}
   k k'=-\frac{1}{2}h'_{ij}k^i k^j \rightarrow k'=-\frac{1}{2}h'_{ij}k n^i n^j
\end{align}
where in the last step we defined the unit vectors $n^i\equiv k^i/k$. On the other hand, from the i-th component of the geodesic equation, these unit vectors evolve as
\begin{align}
 n'^i=\frac{1}{2}h'_{kl}n^k n^l n^i-\frac{1}{2}h'^i_j n^j-\Gamma^{i}_{jk}n^j n^k.   
\end{align}
Now we can write the Boltzmann equation as
\begin{align}
    \frac{d}{d\eta}f (\eta,x^i,k,n^i)\equiv\frac{\partial f }{\partial \eta}+\frac{\partial f}{\partial x^i}x'^i+\frac{\partial f }{\partial k}k'+\frac{\partial f }{\partial n^i}n'^i=a C(\eta,\vec x,\vec k)
\end{align}
On the other hand, we decompose the distribution function into background and perturbations as
\begin{align}
    f(\eta,x^i,k,n^i)=\overline f(\eta,k)(1+\Phi (\eta,x^i,k,n^i)),
\end{align}
where $\overline f(\eta,k)$ is related to the single-particle occupation number, $\overline f(k)=\sum_\lambda n_{k\lambda}$, via Eq.~(\ref{occupationnumberfinal}). Substituting this expression into the left-hand side of the Boltzmann equation and omitting the arguments of the functions for simplicity of notation, we obtain
\begin{align}
 &\frac{d \overline f}{d\eta}(1+\Phi)+\overline f\frac{d\Phi}{d\eta}=\frac{\partial \overline f}{\partial\eta}(1+\Phi)+\overline f\frac{\partial\Phi}{\partial\eta}+\overline f\frac{\partial\Phi}{\partial x^i}n^i-\frac{1}{2}h'_{ij}k n^i n^j \frac{\partial \overline f}{\partial k} = a C(\vec k,\vec x), 
\end{align}
where we used the expression of $k'$ in Eq.~(\ref{kprime}), $x'^i=P^i/P^0\approx n^i$ and consistently expand each term up to first order in perturbations. Finally, we need to expand the collision term into background and perturbations. The energies take the form
\begin{align}
    &\vec p_\phi =m_\phi \vec v_\phi, \quad E_{\phi}=m_\phi\\
    &E_1 = \lvert\vec p_1\rvert\\
    &E_2=\lvert\vec p_2\rvert = \lvert \vec p_\phi-\vec p_1\rvert\approx \lvert\vec p_1\rvert-m_\phi \frac{\vec v_\phi \cdot \vec p_1}{\lvert \vec p_1\rvert}=\lvert\vec p_1\rvert\left(1- \frac{m_\phi   v_\phi \cos\theta  }{\lvert \vec p_1\rvert}\right),
\end{align}
where $\theta$ is the angle between the path of the ALP and one of the dark photons. Then, the collision term is
\begin{align}
    C(\vec p_1)=\frac{(\overline\rho_\phi+\overline P_\phi+\delta\rho_\phi+\delta P_\phi)\lvert\mathcal M\rvert^2 \pi}{8 m_\phi^2\lvert\vec p_1\rvert^2}\left(1+\frac{m_\phi   v_\phi \cos\theta}{\lvert\vec p_1\rvert}\right)\delta\left(\lvert\vec p_1\rvert-\frac{1}{2}m_\phi-\frac{1}{2}m_\phi v_\phi  \cos\theta \right).
\end{align}
Collecting the background term
\begin{align}
    \overline{C}(\vec p_1)=\frac{\overline\rho_\phi \lvert\mathcal M\rvert^2 \pi}{8 m_\phi^2 \lvert\vec p_1\rvert^2}\delta\left(\lvert \vec p_1\rvert-\frac{1}{2}m_\phi\right)
\end{align}
Hence, the background distribution function satisfies
\begin{align}
    \frac{d \overline f}{d\eta}=\frac{\partial \overline f}{\partial \eta}=a \frac{\overline\rho_\phi \lvert\mathcal M\rvert^2 \pi}{8 m_\phi^2 \lvert\vec p_1\rvert^2}\delta\left(\lvert \vec p_1\rvert-\frac{1}{2}m_\phi\right)=a^3\frac{\overline\rho_\phi \lvert\mathcal M\rvert^2 \pi}{8 m_\phi^2 k^2}\delta\left( \frac{k}{a}-\frac{1}{2}m_\phi\right)=a^4\frac{\overline\rho_\phi \lvert\mathcal M\rvert^2 \pi}{8 m_\phi^2 k^2}\delta\left( k-\frac{1}{2} a m_\phi\right)
\end{align}
To obtain the squared of the amplitude, $\lvert\mathcal M\rvert^2$, we first take the derivative of Eq.~(\ref{occupationnumberfinal}) with respect to conformal time
\begin{align}
    n'_{k\lambda}=\pi\frac{g^2_{\phi\rm DR } \varphi_{\rm osc}^2}{32k^2}\frac{a_{\rm osc}^3 m_{\phi}^3}{\mathcal{H}(a_k)}\Theta\left(\frac{2k}{m_\phi}-a_{\rm osc}\right)a\mathcal{H}(a)\delta\left(a-\frac{2k}{m_\phi}\right)
\end{align}
Using the delta function, we can set $\mathcal H(a)=\mathcal H(a_k)$. On the other hand, using the ALP energy density in Eq.~(\ref{ALPenergydensity}), the expression becomes
\begin{align}
\frac{d\overline f}{d\eta}=    \sum_{\lambda=\pm} n'_{k\lambda}=2a^4\overline\rho_\phi \frac{\pi g^2_{\phi \rm DR}}{16k^2}m_\phi \Theta (a-a_{\rm osc})\delta \left(a-\frac{2k}{m_\phi}\right)=a^4\overline\rho_\phi \frac{\pi g^2_{\phi \rm DR}}{16k^2}m^2_\phi \Theta (a-a_{\rm osc})\delta \left(k-\frac{1}{2}a m_\phi\right).
\end{align}
Hence, we obtain
\begin{align}
    \lvert\mathcal M\rvert^2=\frac{g^2_{\phi\rm DR}m_\phi^4}{2}\Theta (a-a_{\rm osc})=32\pi m_\phi\Gamma_\phi\Theta (a-a_{\rm osc}) ,
\end{align}
expression that coincides with the QFT result. On the other hand, the linear term in perturbations of the collision term is
\begin{align}
\delta C(\eta,\vec x,\vec k)
&= \frac{a^3|\mathcal M|^2 \pi}{8m_\phi^2 k^2}
\Bigg[
\left(\delta\rho_\phi+\delta P_\phi
+ a\frac{\overline\rho_\phi m_\phi v_\phi\cos\theta}{k}\right)
\delta\left(k-\frac{1}{2}a m_\phi \right)-
\nonumber\\
&\quad
- \frac{1}{2}a\overline\rho_\phi m_\phi v_\phi \cos\theta\,
\frac{\partial}{\partial k}
\delta\left(k-\frac{1}{2}a m_\phi \right)
\Bigg].
\end{align}
so that the perturbed distribution function $\Phi$ satisfies
\begin{align}
    a\overline C (\eta,k)\Phi + \overline f\frac{\partial\Phi}{\partial\eta}+\overline f\frac{\partial\Phi}{\partial x^i}n^i-\frac{1}{2}h'_{ij}k n^i n^j=a\delta C(\eta,\vec k,\vec x),
\end{align}
where we used the Boltzmann equation for the background distribution function.

Now we relate the components of the energy momentum tensor of radiation to the distribution function using Eq.~(\ref{energymomentumtensorDR}) and the definitions of $k$ and $n^i$. At linear order in perturbations
\begin{align}
    &(\overline T_{\rm DR})^0_0+(\delta T_{\rm DR})^0_0=a^{-4}\int\frac{dk~d\Omega}{(2\pi)^3}k^3 \overline{f}+a^{-4}\int\frac{dk~d\Omega}{(2\pi)^3}k^3 \overline{f}\Phi\\
 &(\overline T_{\rm DR})^i_j+(\delta T_{\rm DR})^i_j=-a^{-4}\int\frac{dk~d\Omega}{(2\pi)^3}k^3 n^i n_j\overline{f}-a^{-4}\int\frac{dk~d\Omega}{(2\pi)^3}k^3 n^i n_j \overline{f}\Phi \\
     &(\delta T_{\rm DR})^0_i=a^{-4}\int\frac{dk~d\Omega}{(2\pi)^3}k^3 n_i \overline{f}\Phi.
\end{align}
Evaluating $\nabla_\mu \delta (T_{\rm DR})^{u}_\nu$ we get
\begin{align}\label{nablaTradperturbations}
&\nabla_\mu \delta (T_{\rm DR})^{u}_0=a\Gamma_\phi(\delta\rho_\phi+\delta P_\phi)\Theta(a-a_{\rm osc})\\
&\nabla_\mu \delta (T_{\rm DR})^{u}_i=-\Gamma_\phi\overline\rho_\phi a v_\phi \delta_i^3
\Theta(a-a_{\rm osc}),
\end{align}
where we fixed the z axis as the direction of the motion of the ALP, we used the Christoffel symbols in Eq.~(\ref{perturbedchristoffels}), integration by parts, the Boltzmann equation, and the properties
\begin{align}
& P_i=g_{ij} P^j=-\left(\delta_{ij}+\frac{1}{2}h_{ij}\right)k^j,\quad \left| g \right| =a^8(1+h),\quad d^3P=\left| \frac{\partial P_i}{\partial k^j}\right| d^3k=\left(1+\frac{1}{2}h\right)d^3k\\
    &\int d\Omega n^i=0,\quad \int d\Omega n^i n_j=\frac{2}{3}\delta^i_j,\quad \int d\Omega n^i n^j n^k=0.
\end{align}
On the other hand, the perturbed energy-momentum tensor of the dark radiation fluid is given by
\begin{align}
    (T_{\rm DR})^{\mu}_\nu=\begin{pmatrix}
      \overline{\rho}_{\rm DR}+\delta\rho_{\rm DR} & -(\overline{\rho}_{\rm DR}+\overline{p}_{\rm DR}) (v_{\rm DR})_i\\
      \left(\overline{\rho}_{\rm DR}+\overline{p}_{\rm DR}\right)v_{\rm DR}^i &-(\overline{P}_{\rm DR}+\delta P_{\rm DR})\delta^i_j-\Pi^i_j
    \end{pmatrix}.   
\end{align}
Dark radiation is a perfect fluid at both levels, background and perturbations. Hence $\delta P_{\rm DR}=1/3\delta\rho_{\rm DR}$. If we compute $\nabla_{\mu} (\delta T_{\rm DR})^{\mu}_\nu$, and compare with Eq.~(\ref{nablaTradperturbations}), we obtain
\begin{align}
 &\delta'_{\rm DR}=-\frac{4}{3}\left(\theta_{\rm DR}+\frac{1}{2}h'\right)+a\Gamma_\phi \frac{\overline{\rho}_\phi}{\overline\rho_{\rm DR}}\left[\left(1+\frac{\delta P_\phi}{\delta\rho_\phi}\right)\delta_\phi-\delta_{\rm DR}\right]\\
 & \theta'_{\rm DR}=-\frac{1}{4}\nabla^2\delta_{\rm DR}-\frac{3}{4\overline\rho_{\rm DR}}\partial_i\partial_j \Pi^{ij}-a\Gamma_\phi\frac{3}{4 \overline\rho_{\rm DR}}\left(\frac{4}{3}\theta_{\rm DR}-\theta_{\phi}\right)
\end{align}
where we used $\Pi^i_i=0$ and $\theta_{\rm DR}=\partial_i v^i_{\rm DR}$. These equations are valid for $a>a_{\rm osc}$, where DR is produced. In Fourier space, the shear can be written as
\begin{align}
    \Pi^{ij}(\vec x)=-\int\frac{d^3q}{(2\pi)^3}e^{i\vec q\cdot \vec x}\left(\frac{q^i q^j}{q^2}-\frac{1}{3}\delta^{ij}\right)\frac{3}{2}(\overline\rho_{\rm DR}+\overline P_{\rm DR})\sigma_{\rm DR},
\end{align}
where $\sigma_{\rm DR}$ is the scalar shear, that has no counterpart defined in real space. Thus
\begin{align}
    \partial_i\partial_j\Pi^{ij}\rightarrow  q^2 \frac{4}{3}\overline \rho_{\rm DR}\sigma_{\rm DR}
\end{align}
and the equations in Fourier space become
\begin{align}
 &\widetilde \delta'_{\rm DR}=-\frac{4}{3}\left(\widetilde \theta_{\rm DR}+\frac{1}{2}h'\right)+a\Gamma_\phi \frac{\overline{\rho}_\phi}{\overline\rho_{\rm DR}}\left[\left(1+\frac{\widetilde{\delta P}_\phi}{\widetilde{\delta\rho}_\phi}\right)\widetilde \delta_\phi-\widetilde \delta_{\rm DR}\right]\\
 & \theta'_{\rm DR}=\frac{1}{4}q^2\widetilde \delta_{\rm DR}-q^2\sigma_{\rm DR}-a\Gamma_\phi\frac{3}{4 \overline\rho_{\rm DR}}\left(\frac{4}{3}\widetilde \theta_{\rm DR}-\widetilde \theta_{\phi}\right).
\end{align}
The previous equation depends on $\sigma_{\rm DR}$, defined as the second multipole of the Boltzmann hierarchy associated with the integrated phase-space distribution function $F$, given by
\begin{align}\label{integratedF}
 F=\frac{\int dk~ k^3 \overline f \tilde{\Phi} }{\int dk~k^3 \overline f}r_{\rm DR};\quad r_{\rm DR}=\frac{\overline\rho_{\rm DR}a^4}{\rho^0_{\rm cr}}.   
\end{align}
Here, $\rho^0_{\rm cr}$ is the present-day critical energy density of the Universe, and $\tilde\Phi(\eta,\vec q,\vec k)$ denotes the Fourier transform of the linear perturbation of the distribution function $\Phi(\eta,\vec x,\vec k)$,
 which satisfies the Boltzmann equation in Fourier space
\begin{align}\label{fouriereqforphi}
  a\overline C (k)\tilde{\Phi} + \overline f\frac{\partial\tilde \Phi}{\partial\eta}+i\overline f  q_i n^i\tilde{\Phi}-\frac{1}{2} \left[\frac{q_i q_j}{q^2} h'+\left(\frac{q_i q_j}{q^2}-\frac{1}{3}\delta_{ij}\right)6\eta\right]k n^i n^j=a\widetilde{\delta C}(\eta,\vec q,\vec k),    
\end{align}
with
\begin{align}
\widetilde{\delta C}(\eta,\vec q,\vec k)
&= \frac{a^3|\mathcal M|^2 \pi}{8m_\phi^2 k^2}
\Bigg[
\left(\widetilde{\delta\rho}_\phi+\widetilde{\delta P}_\phi
+ a\frac{\overline\rho_\phi m_\phi \widetilde v_\phi\cos\theta}{k}\right)
\delta\left(k-\frac{1}{2}a m_\phi \right)-
\nonumber\\
&\quad
- \frac{1}{2}a\overline\rho_\phi m_\phi \widetilde v_\phi\cos\theta\,
\frac{\partial}{\partial k}
\delta\left(k-\frac{1}{2}a m_\phi \right)
\Bigg].
\end{align}
where $v^i_\phi(\vec x)$ is the velocity field, and $\widetilde 
v_\phi$ denotes its Fourier transform.
We further assume that $v^i_\phi$ is irrotational, $\nabla\times \vec v_\phi=0$, which in Fourier space implies that $\vec {\tilde v}(\vec q)_\phi\parallel  \vec q$. As a result, that $q_i n^i =q \hat q^i n_i=q\hat v_\phi^i n_i=q\cos\theta$, where $\theta$ is the angle between the direction of propagation of a dark photon and the ALP velocity. Then
\begin{align}
 a\overline C (k)\widetilde{\Phi} + \overline f\frac{\partial\tilde \Phi}{\partial\eta}+i\overline f  q \cos\theta\widetilde{\Phi}-\frac{1}{2}k\left[  h'\cos^2\theta+\left(\cos^2\theta-\frac{1}{3}\right)6\eta\right]=a\widetilde{\delta C}(\eta,\vec q,\vec k).   
\end{align}
We now differentiate Eq.~(\ref{integratedF}) with respect to conformal time, using the background evolution of $\overline\rho_{\rm DR}$. First
\begin{align}
    r'_{\rm DR}=a\Gamma_\phi\overline\rho_\phi \frac{r_{\rm DR}}{\overline\rho_{\rm DR}}.
\end{align}
Then, by using Eq.~(\ref{fouriereqforphi}), we find that
\begin{align}
 F'=r'_{\rm DR}\left[\left(1 +\frac{\widetilde{\delta P}_\phi}{\widetilde {\delta\rho}_\phi}\right)\widetilde \delta_\phi+3\widetilde v_\phi\cos\theta\right]-2\left[h'\cos^2\theta+\left(\cos^2\theta-\frac{1}{3}\right)6\eta' \right]r_{\rm DR}-iq\cos\theta F.   
\end{align}
Assuming azimuthal symmetry, $F$ can be expanded in Legendre polynomials as
\begin{align}
    F=\sum_{\ell=0}^{\infty}(-i)^\ell (2\ell+1)F_\ell P_\ell(\cos\theta).
\end{align}
The Legendre polynomials satisfy
\begin{align}
    x P_\ell(x)=\frac{(\ell+1)P_{\ell+1}(x)+\ell P_{\ell-1}(x)}{2\ell+1},\quad \int_{-1}^1
dx P_{\ell}(x) P_{\ell'}(x)=\frac{2}{2\ell+1}\delta_{\ell\ell'}.
\end{align}
The first Legendre polynomials are
\begin{align}
    P_0(x)=1,\quad P_1(x)=x,\quad  P_2(x)=\frac{1}{2}(3\cos^2 x-1).
\end{align}
Hence, the Boltzmann hierarchy reads
\begin{align}
    &F_0'=r'_{\rm DR}\left(1+\frac{\widetilde{\delta P}_\phi}{\widetilde{\delta\rho}_\phi}\right)\widetilde \delta_\phi-\frac{2}{3}h' r_{\rm DR}-q F_1\\
    &F_1'=r'_{\rm DR}\frac{\widetilde \theta_{\rm DR}}{q}+q\frac{F_0}{3}-\frac{2q}{3}F_2\\
    &F_2'=\frac{4}{15}(h'+6\eta')r_{\rm DR}+q\frac{F_1}{5}-q\frac{3}{5}F_3\\
    &F'_{\ell}=\frac{q}{2\ell+1}\left(\ell F_{\ell-1}-(\ell+1)F_{\ell+1}\right),\quad \ell \geq 3.
\end{align}
Finally, the relation between the multipoles $F_{\ell}$ and the perturbations $\widetilde{\delta}_{\rm DR}$, $\widetilde{\theta}_{\rm DR}$, and $\sigma_{\rm DR}$ follows from the definition of $F$, together with the expression for the energy–momentum tensor and the orthogonality of the Legendre polynomials
\begin{align}
 &F_0=a^4\Omega_{\rm DR} \widetilde{\delta}_{\rm DR}\\
 &F_1=\frac{4}{3}a^4\Omega_{\rm DR}\frac{\widetilde\theta_{\rm DR}}{q}\\
 &F_2=2a^4 \Omega_{\rm DR}\sigma_{\rm DR},
\end{align}
where $\Omega_{\rm DR}=\overline\rho_{\rm DR}/\rho^0_{\rm cr}$.

\subsubsection{Metric perturbations}

Finally, we write the equations for the metric perturbations. Following Ref.~\cite{Ma:1995ey}
\begin{align}
 &\frac{1}{2}\mathcal{H}h'-q^2\upsilon=a\elev2\frac{\widetilde{\delta\rho}}{2M_p\elev2}\\
 &q^2\upsilon'=a\elev2\frac{(\overline{\rho}+\overline{p})\widetilde{\theta}}{2M_p\elev2}\\
 &h''+2\mathcal H h'-2q^2\upsilon =-3a^2\frac{ \widetilde{\delta p}}{M_p\elev2}\\
 &h''+6\upsilon''+2\mathcal{H}\left(h'+6\upsilon'\right)-2q\elev2\upsilon=-3 a\elev 2\frac{\left(\overline\rho+\overline p\right)\sigma}{M_p\elev2}
\end{align}
where $\overline\rho$, $\overline p$ are the total background density and pressure; while $\widetilde{\delta\rho}$, $\widetilde{\delta p}$, $\widetilde \theta$ and $\sigma$ are the total perturbation in density, pressure, velocity and shear. The last equation was simplified using the background Friedmann equations (\ref{friedmanneqconformal}). This completes the full set of equations that are implemented in CLASS \cite{Diego_Blas_2011}.

\section{Numerical analysis}\label{Bayesiananalysis}
\subsection{Constraints on the ALP model parameters}
Let us first fix the parameter space we will explore in our numerical analysis:
\begin{itemize}
    \item [1.] We assume that the ALP starts oscillating immediately after it gets trapped by one of the cosine potential minima. This condition translates into
    \begin{align}
        m_\phi>3H_{\rm osc}=3H_0\sqrt{\Omega^0_{\rm rad}}a_{\rm osc}^{-2},
    \end{align}
    since DM must be generated during radiation domination epoch.
    \item [2.] Since the ALP is our DM candidate, its energy density at $a=a_{\rm osc}$ is matched to
    \begin{align}
        \rho_\phi(a_{\rm osc})=\Omega\elev{\rm CDM}_{\rm ini}3 H\elev 2_0 M_p\elev 2 a_{\rm osc}\elev{-3},
    \end{align}
    where $\Omega\elev{\rm CDM}_{\rm ini}$ is a CLASS input parameter that describes the dark matter evolution assuming it does not decay. The physical value is then obtained by correcting this parameter using a shooting method that includes the effect of dark matter decay. On the other hand, since the ALP energy density when it gets trapped by the cosine potential is $\rho_\phi=m_\phi\elev 2 f_\phi\elev 2$, one gets a relation between the $m_\phi$, $f_{\phi}$ and $a_{\rm osc}$
    \begin{align}\label{relation mfa_osc}
      m_\phi\elev 2f_\phi\elev 2 =\Omega\elev{\rm CDM}_{\rm ini}3 H\elev 2_0 M_p\elev 2 a_{\rm osc}\elev{-3}. 
    \end{align}

    \item [3.] CLASS starts iterating at the scale factor $a_{\rm ini}=10^{-14}$. For us, this imposes a lower bound on $a_{\rm tr}$, so that the phase with constant $\omega_\phi$ has an impact on the Universe history. Hence $a_{\rm tr}>10^{-14}$ or $a_{\rm osc}>10\elev{-14}(2r_{\phi})\elev{1/6}$. 

 \item [4.] DM must exist before recombination. Hence, we impose an upper bound on $a_{\rm osc}$, so that $a_{\rm osc}<10^{-4}$.
    \item [5.] To preserve the standard history of the Universe, meaning that, an epoch of radiation domination is followed by a matter dominated epoch, we impose that the scalar field does not dominate the energy content of the Universe at $a=a_{\rm ini}$ for any value of $\omega_{\phi}$. In CLASS, the default initial energy density of any component at $a = a_{\rm ini}$ is limited to $10^{-4}\,\rho_{\rm rad}$. We relax this bound to $10^{-2}\,\rho_{\rm rad}$, thus
    \begin{align}
        \Omega_{\rm ini}^{\rm CDM}a_{\rm osc}^{3\omega_{\phi}}(2r_{\phi})^{\frac{1-\omega_{\phi}}{2}}a_{\rm ini}^{-3(1+\omega_{\phi})}<10^{-2}\Omega^0_{\rm rad}a_{\rm ini}^{-4}.
    \end{align}
Combining (3), (4) and (5) and standard values $\Omega_{\rm ini}^{\rm CDM}=0.3$, $\Omega^0_{\rm rad}=9 \times 10^{-5}$ (see for instance \cite{Baumann:2022mni}), we get an upper bound for $r_{\phi}$ 
\begin{align}
    \log_{10}r_{\phi}<38+\log_{10}\left(\frac{3}{2}\right)\approx 38.18. 
\end{align}
       \item [6.] From Eq.~(\ref{relation mfa_osc}), it appears that we can only constrain the product $m_\phi f_\phi$. However, the parameter $c_1$ in the perturbations (see Eq.~(\ref{scalar_field_sound_speed})) depends only on the ALP mass. Therefore, our analysis can indeed place bounds on both parameters. To derive an upper bound on the scale $f_\phi$, we combine conditions (1) and (2) to obtain
    \begin{align}\label{upperf_phi}
     \frac{f_\phi}{M_p}<\sqrt{\frac{1}{3}\frac{\Omega_{\rm ini}^{\rm CDM}}{\Omega^0_{\rm rad}}a_{\rm osc}}.   
    \end{align}
For the numerical analysis, we will set the lower bound $f_\phi\geq 1\, \rm TeV$.
\item [7.] We only want to produce DR from recombination onward since the CMB constrains the amount of DM at that moment. Then, we impose an upper bound on the decay width of the scalar field $\Gamma_\phi\leq 10^7~\rm{Km } ~\rm s^{-1} Mpc^{-1}$. 

\end{itemize}

\subsection{Datasets}
To perform our MCMC analysis, we use Cobaya \cite{Torrado:2020dgo}. The likelihoods included in the global fit are the following:
\begin{itemize}
    \item [1.] Low-$\ell$ TT power spectra from the \textit{Planck} 2018 likelihood \cite{Planck:2019nip}.
    \item [2.] LoLLiPoP: Low-$\ell$ Likelihood Polarized for Planck \cite{Tristram:2020wbi}.
    \item [3.] HiLLiPoP: High-$\ell$ Likelihood Polarized for Planck \cite{Tristram:2023haj}.
     \item [4.] Planck PR4 (NPIPE) lensing likelihoods \cite{Carron:2022eyg,Carron:2022eum}. It contains the Planck lensing and ISW-lensing likelihoods built from Planck PR4(NPIPE) data.
    \item [5.] DES‑Dovekie sample \cite{DES:2025sig}: updates the DES‑SN 5YR release \cite{DES:2024jxu} with improved photometry, low‑z cross-calibration, retrained SALT3 light curves, and a corrected host-galaxy color law. It includes SN over the redshifts range $0.025 < z < 1.13$.
    \item [6.] Pantheon+ sample \cite{Brout:2022vxf}. Provides luminosity distance measurements of SN over the redshift range $0.001<z<2.26$.
    \item [7.] BAO data from DESI \cite{DESI:2024mwx,DESI:2024uvr,DESI:2024lzq}. It includes measurements of galaxy and quasar baryon acoustic oscillation over the redshift range $0.1<z<4.1$. 
    \item [8.] A Gaussian prior on $H_0$ to take into account the results of the SH0ES~\cite{Riess:2021jrx} collaboration.

\end{itemize}

\begin{figure}[ht!]
    \centering
    \subfigure[]{\includegraphics[height=6cm]{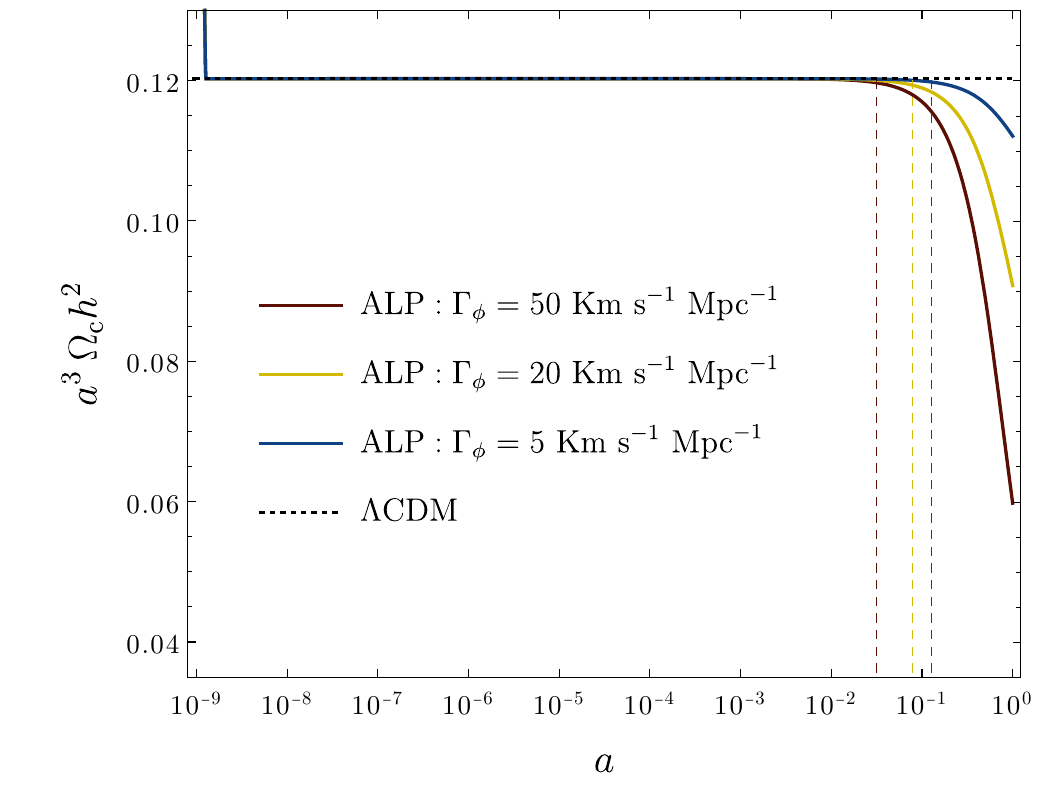}}\hfill
      \subfigure[]{\includegraphics[height=6cm]{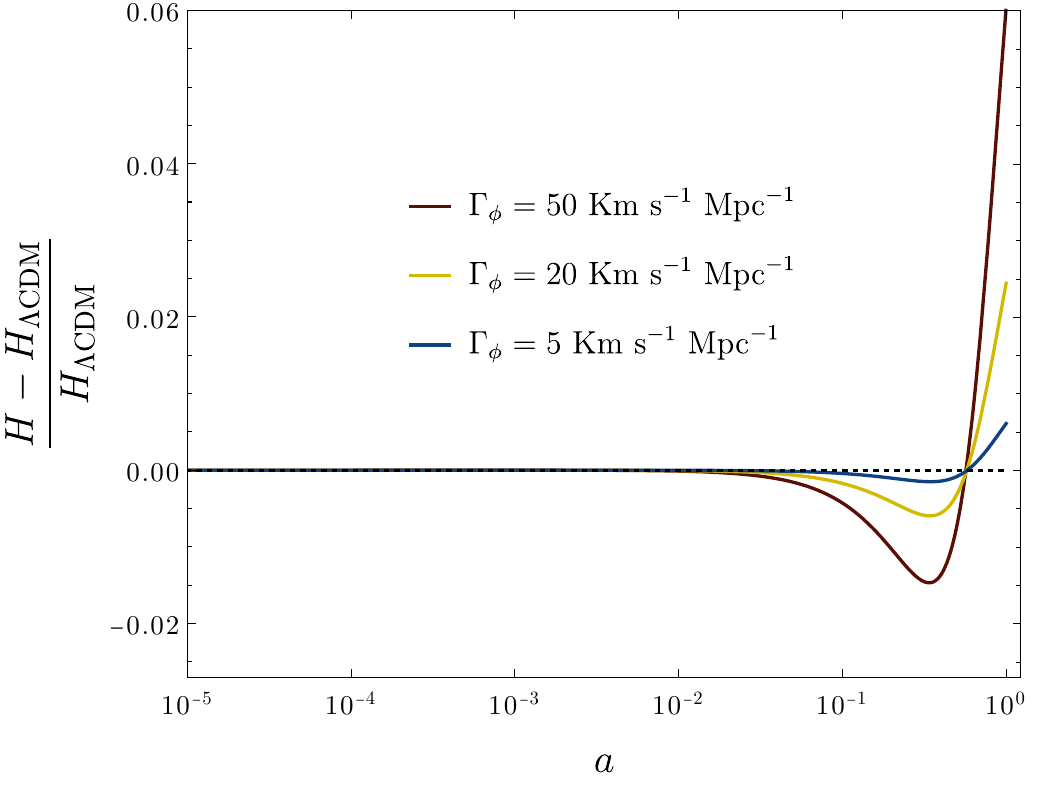}}
    \caption{\textit{Left}: Comparison between the evolution of DM in the ALP model for a selected values of $\Gamma_\phi$ and $\Lambda\rm CDM$. The vertical dashed lines represent the moment when the ALP coupling to  DR becomes effective.
    \textit{Right:} Hubble parameter as a function of the scale factor in the ALP model for the selected values of $\Gamma_\phi$. The Hubble parameter is obtained using the `shooting' method implemented in CLASS. In both cases, the rest of the parameters are fixed to those in Table~\ref{tab:base_lcdm_params}.}
\label{fig:hubble_densities}
\end{figure}

\subsection{Results}

 \begin{table*}[ht!]
 \centering 
\begin{tabular} { |l c|}\hline
 Parameter & $\Lambda$CDM\\
\hline\hline
{$\log(10^{10}A_\mathrm{s})$}& $3.0447838$\\

{$n_\mathrm{s}   $} & $  0.9660499      $\\

{$100\Omega_\mathrm{b} h^2$} & $ 2.238280$\\

{$\Omega_\mathrm{c} h^2$} &  $0.1201075 $\\

{$\tau_\mathrm{reio}$} & $ 0.054308$\\

{$100\theta_\mathrm{s}$} &  $1.041783 $\\
\hline
\end{tabular}
\caption{Baseline $\Lambda$CDM parameters used in the computation of the benchmark points, as specified in the CLASS \textit{.ini} file.}\label{tab:base_lcdm_params}
\end{table*}

 \begin{table*}[ht!]
 \centering
\begin{tabular} { |l| c c|}\hline
 Parameter &  ALP model&$\Lambda$CDM\\
\hline\hline
{$\log(10^{10}A_\mathrm{s})$}& $3.053\pm 0.012 $&
$ 3.053\pm 0.012$\\

{$n_\mathrm{s}   $} & $  0.9711\pm 0.0035      $& $0.9709\pm 0.0034  $\\

{$100\Omega_\mathrm{b} h^2$} & $2.236^{+0.012}_{-0.011}$& $ 2.237^{+0.012}_{-0.010}     $\\

{$\Omega_\mathrm{c} h^2$} &  $0.11747\pm 0.00074 $& $0.11750\pm 0.00074 $\\

{$\tau_\mathrm{reio}$} & $0.0625\pm 0.0062$ & $0.0631\pm 0.0060 $\\

{$100\theta_\mathrm{s}$} &  $1.04195^{+0.00028}_{-0.00023} $& $ 1.04197\pm 0.00023 $\\

{$\omega_{\phi}          $}  &
$< -0.0370  $&
---                         \\
{$\log_{10}r_{\phi} $} & 
$< 11.1  $&
---  \\
{$\log_{10}(f_\phi/M_p)$} & 
$-9.4^{+3.9}_{-4.8} $&
---  \\
{$\log_{10}a_{\rm osc}$} & 
$ -8.9^{+3.2}_{-2.6} $&
---  \\
{$\log_{10}\Gamma_{\phi}$} & 
$ < -1.88   $&
---  \\

\hline\hline

$\Omega_\mathrm{m}         $ &$ 0.3006\pm 0.0045$& $ 0.3013\pm 0.0043$\\

$S_8$&$0.8133^{+0.0051}_{-0.013} $& $ 0.8099\pm 0.0083$\\

$H_0$ & $68.32\pm 0.33 $& $   68.29\pm 0.33         $\\
\hline
\end{tabular}

\caption{The 68\% confidence limits on the base $\Lambda$CDM and ALP model parameters, along with the derived parameters $\Omega_m$, $S_8$ and $H_0$, for the ALP model and $\Lambda$CDM analysis. The units of $H_0$ and $\Gamma_{\phi}$ are $\rm{Km/s/Mpc }$.}\label{tab:68_confidence_intervals}
\end{table*}

\begin{figure}[h]
    \centering
{\includegraphics[height=6cm]{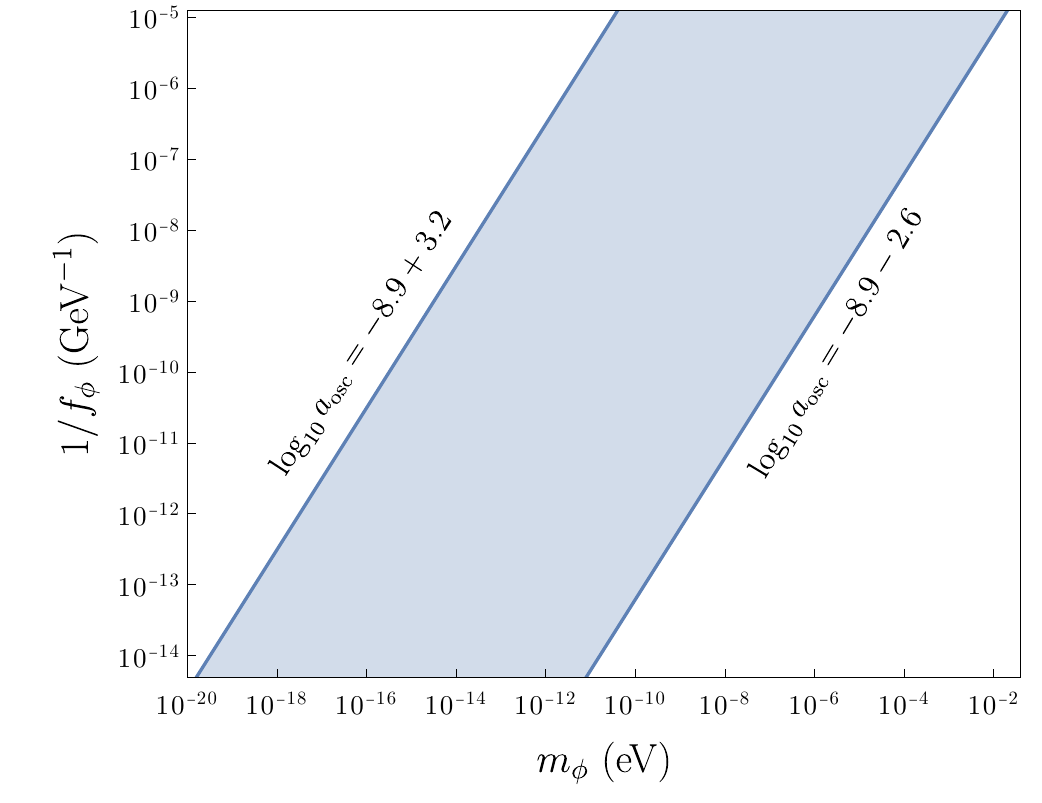}}\hfill
        \subfigure[]{\includegraphics[height=6cm]{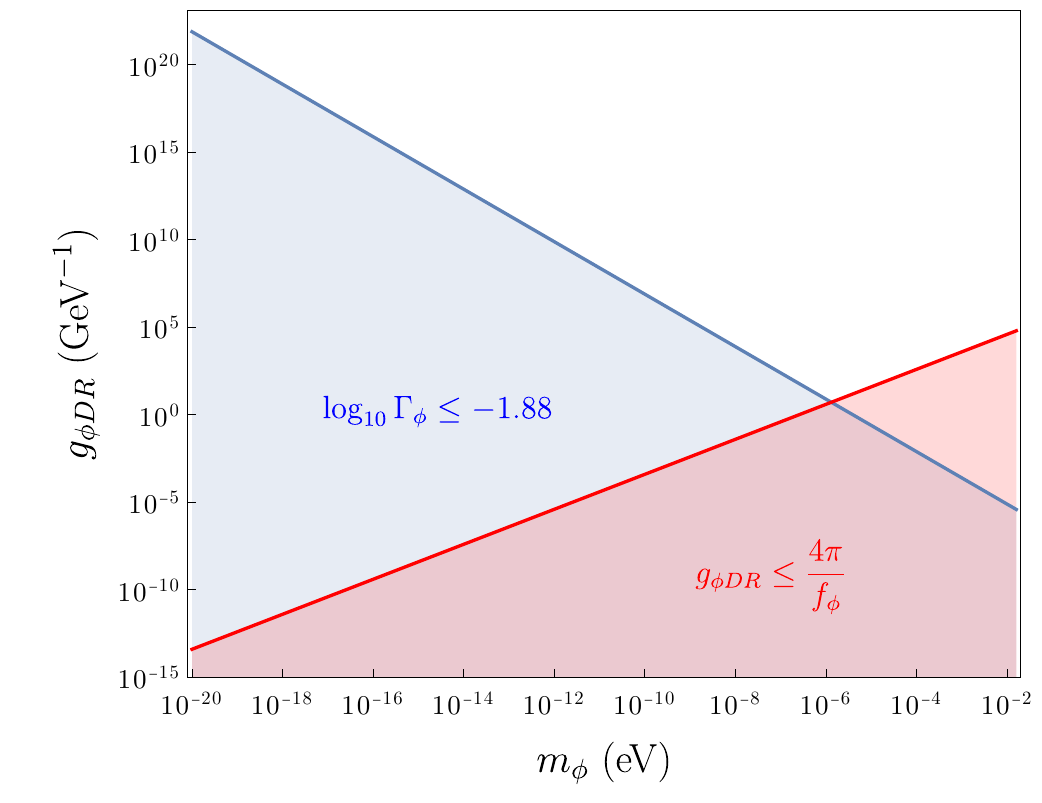}}
    \caption{{\it{Left}}: Allowed region in the $m_\phi$-- $f_\phi$ plane consistent with Table~\ref{tab:68_confidence_intervals}. {\it Right}: In blue, the allowed region in the $m_\phi$--$g_{\phi\mathrm{DR}}$ plane from Table~\ref{tab:68_confidence_intervals}; in red, the perturbativity bound. For $m_\phi \gtrsim 10^{-6}~\mathrm{eV}$, cosmological constraints are more stringent than the perturbativity bound.
    }\label{fig:results}
\end{figure}

\begin{figure}[ht!]
\includegraphics[width=19cm]{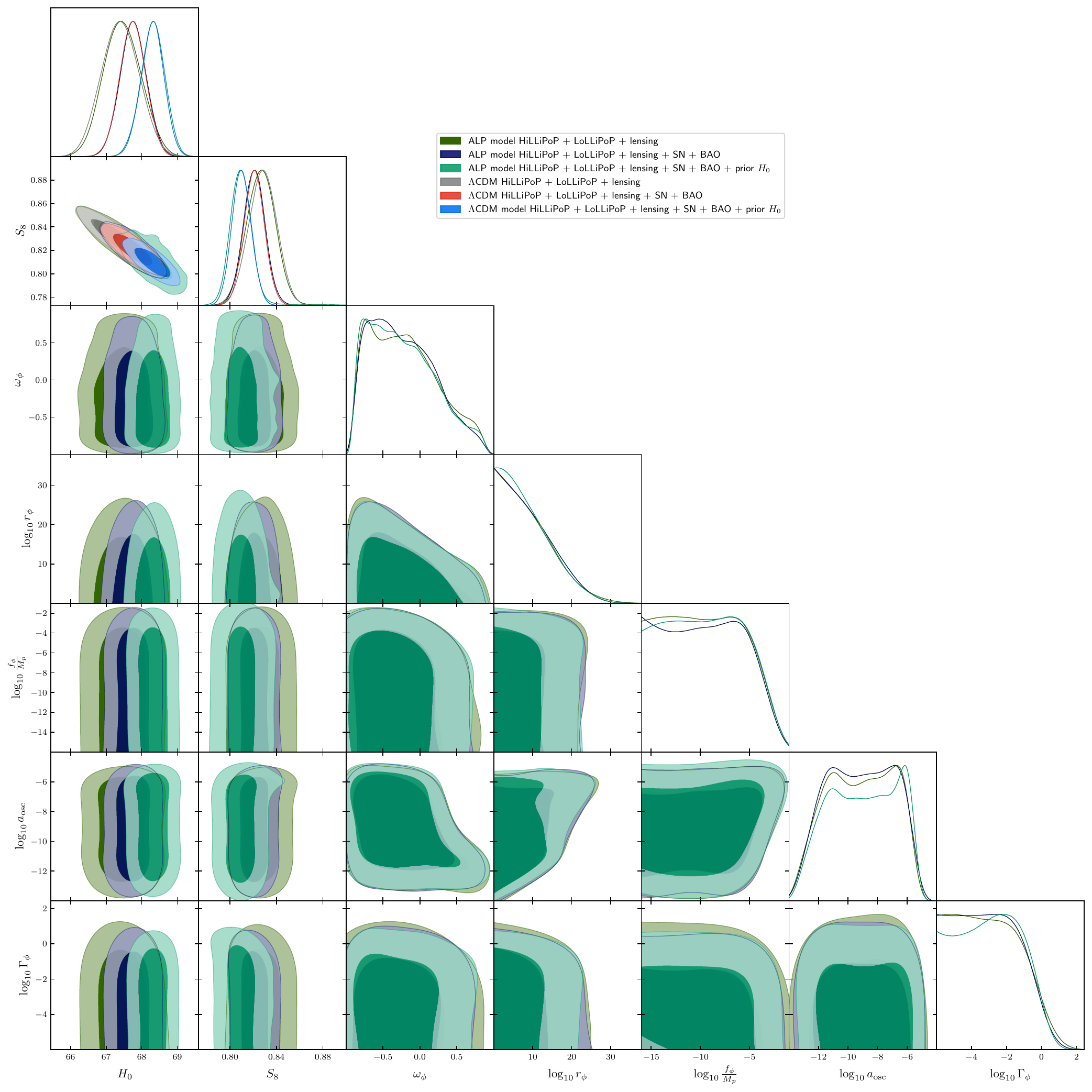}
\centering
\caption{Posteriors on $H_0$ and $S_8$, as well as the $\omega_{\phi}$, $\log_{10}r_{\phi}$, $\log_{10}f_\phi/M_P$, $\log_{10}a_{\rm osc}$ and $\log_{10} \Gamma_{\phi}$ parameters for ALP decaying model analysis.} 
\label{fig:posteriors}
\end{figure}

We run our chains until they are converged with a Gelman-Rubin convergence criterion of $R−1 < 0.06$. The posterior distributions of the parameters $H_0$, $S_8$, $\omega_\phi$, $\log_{10}r_\phi$, $\log_{10}(f_\phi/M_p)$, $\log_{10}a_{\rm osc}$ and $\log_{10}\Gamma_\phi$ are shown in Fig.~\ref{fig:posteriors}. In view of the $68\%$ confidence limits on the ALP model parameters in Table~\ref{tab:68_confidence_intervals}, there is a preference for a negative ALP equation of state, $\omega_{\phi}$, before the transition in the potential. This implies that the ALP potential energy dominates over the kinetic energy, thereby avoiding any possibility of an early phase of kinetic misalignment. 

The parameter $r_{\phi}$, defined as the ratio of the ALP energy density to the height of the cosine potential, along with $a_{\rm osc}$, determine the duration of the phase with constant equation of state and the subsequent kinetic misalignment phase. To provide an estimation for the duration of these phases, we first calculate $a_{\rm tr}$ through Eq.~(\ref{a2})
\begin{align}
    a_{\rm tr}=a_{\rm osc}\left(\frac{1}{2r_{\phi}}\right)^{\frac{1}{6}}\geq 4\times 10^{-14},
\end{align}
where we took the lower bound of $\log_{10}a_{\rm osc}>-8.9-2.6=-11.5$. Using Eq.~(\ref{tinraddom}), valid for a radiation dominated Universe, we find
\begin{equation}
    \Delta t(\omega_{\phi}=\textrm{ const})\approx\frac{1}{2H_0\sqrt{\Omega^0_{\rm rad}}}a_{\rm tr}^2\geq 3.5\times 10^{-8}\, \rm s,
\end{equation}
after the electroweak phase transition. To place an upper bound on the kinetic misalignment phase
\begin{equation}
    \Delta t(\textrm{kin-mis})=\frac{1}{2H_0\sqrt{\Omega^0_{\rm rad}}}(a_{\rm osc}^2-a_{\rm tr}^2)\leq 9.3\times 10^7\, \rm s\approx 1000~yrs,
\end{equation}
where in this case we use $\log_{10}a_{\rm osc}\leq -8.9+3.2=-5.7$.

The global fit places lower and upper bounds on the energy scale $f_\phi$, namely $f_\phi\in \left[80,1.5\times 10^{10}\right]~\rm TeV$, spanning 9 orders of magnitude. Since we work in the pre-inflationary scenario, these bounds also imply an upper bound on the energy scale of inflation, as required by the condition $f_\phi>H_I$. These constraints on $f_\phi$ translate into bounds on the ALP mass via Eq.~(\ref{relation mfa_osc})

\begin{equation}\label{eq:m-f}
   \log_{10}\left(\frac{M_p}{f_\phi}\right) =\log_{10}\left(\frac{m_\phi}{1\,\textrm{eV}}\right)-\frac{1}{2}\log_{10}\left(\Omega_c h^2\right)+\frac{3}{2}\log_{10}a_{\rm osc}+32.67,
\end{equation}
along with the bounds on $\Omega_ch^2$ and $a_{\rm osc}$ in Table~\ref{tab:68_confidence_intervals}, yielding the band in the $(m_\phi,f_\phi)$ plane shown in the first panel of Fig.~\ref{fig:results}. Within the  allowed range of $f_\phi$, the corresponding ALP masses lie in the range $m_\phi\in [10^{-20},10^{-2}]~\rm eV$. It is also worth noting that, compared to the models shown in Fig.~\ref{fig:misalignment_mech}, this model allows for smaller values of $f_\phi$ for a given mass than in the standard misalignment scenario.

Finally, Cobaya constrains the ALP decay width to be $\log_{10}\Gamma_{\phi}\leq -1.88$. While this is not sufficient by itself to resolve the cosmological tensions, Eq.~(\ref{ALPdecaywidth}) allows us to derive bounds for the ALP coupling to DR, $g_{\phi\rm DR}$, as a function of the ALP mass
\begin{equation}
    \log_{10}\left(\frac{g_{\phi\rm DR}}{1\, \rm GeV^{-1}}\right)\leq -8.12-\frac{3}{2}\log_{10}\left(\frac{m_\phi}{1\,\rm eV}\right).
\end{equation}
However, perturbativity imposes an additional constraint,
\begin{equation}\label{eq:g_m}
    g_{\phi\rm DR}=\frac{c}{f_\phi}\leq \frac{4\pi}{f_\phi},
\end{equation}
where $c$ is a dimensionless order 1 parameter bounded by $c\leq 4\pi$. From Eq.~(\ref{eq:m-f}), we can express the scale $f_\phi$ in terms of the ALP mass. This yields
\begin{equation}\label{eq:g_pert}
    \log_{10}\left(\frac{g_{\phi\rm DR}}{1\,\rm GeV^{-1}}\right)\leq\log_{10}\left(\frac{m_\phi}{1\,\rm eV}\right)+6.59.
\end{equation}
In the above expression, we used $\log_{10} a_{\rm osc} \leq -8.9+3.2=-5.7$ from Table~\ref{tab:68_confidence_intervals}. In the second panel of Fig.~\ref{fig:results}, we show the bounds derived from Eqs.~(\ref{eq:g_m}) and (\ref{eq:g_pert}). This plot indicates that, for small ALP masses, the bounds from perturbativity are more stringent than those from cosmology. However, for ALP masses larger than $1\,\mathrm{\mu eV}$, the cosmological bound becomes stronger than perturbativity.

\begin{figure}[ht!]
\centering
\subfigure[]{\includegraphics[height=5.4cm]{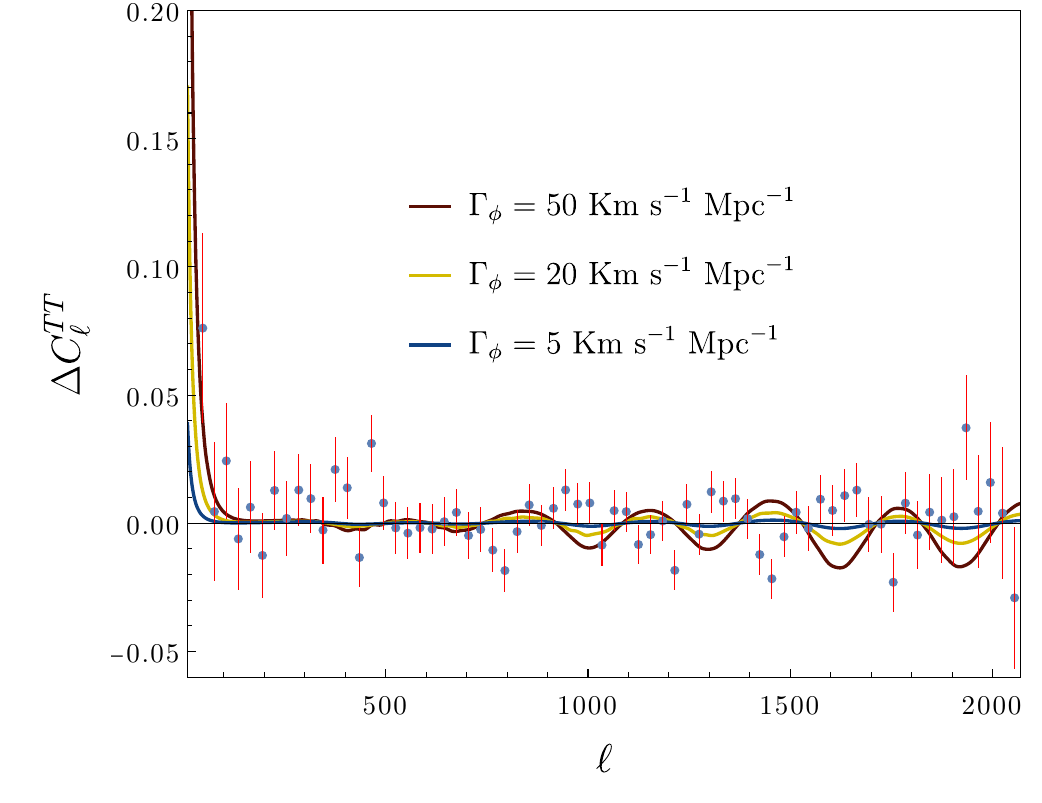}}\hfill
\subfigure[]{\includegraphics[height=5.4cm]{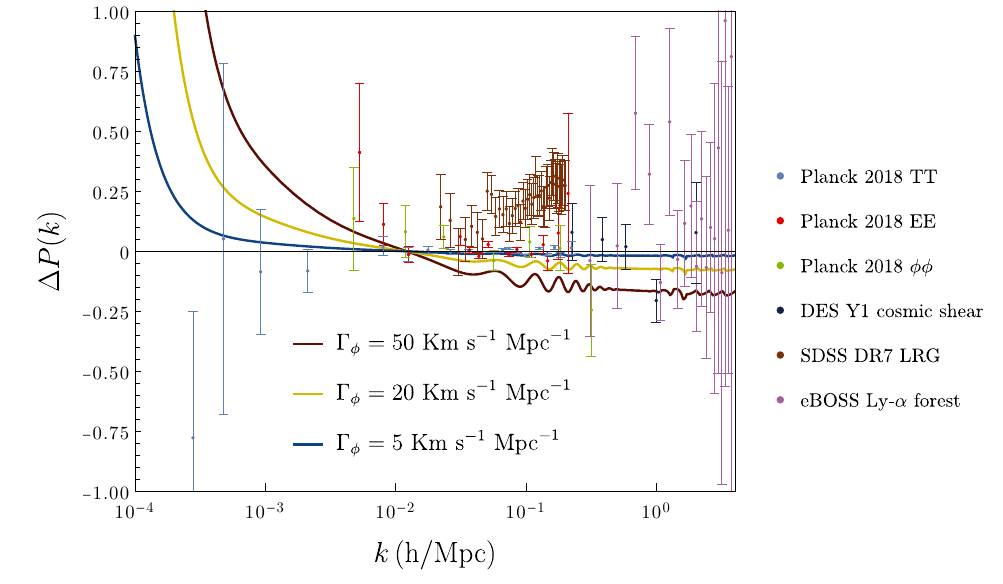}}
\caption{{\it Left:} Induced changes in the CMB temperature power spectrum in the ALP model for selected values of $\Gamma_\phi$. The rest of the parameters are fixed to those in Table~\ref{tab:base_lcdm_params} and $\omega_\phi=-0.037$, $\log_{10}r_\phi=11.1$, $\log_{10}(f_\phi/M_P)=-9.4$, $\log_{10}a_{\rm osc}=-8.9$. 
{\it Right:} Same for the linear matter power spectrum $P (k)$. In each case we define the fractional change $\Delta X\equiv \frac{X-X_{\Lambda\rm{CDM}}}{X_{\Lambda\rm{CDM}}}$, where $X$ is either $C^{TT}_{\ell}$ or $P(k)$. 
The CMB power spectrum data points are extracted from \url{http://pla.esac.esa.int/pla} and for the matter power spectrum from  \url{https://github.com/marius311/mpk_compilation}.
}
\label{fig:perturbations_benchmark}
\end{figure}

\begin{figure}[ht!]
\centering
\subfigure[]{\includegraphics[height=5.4cm]{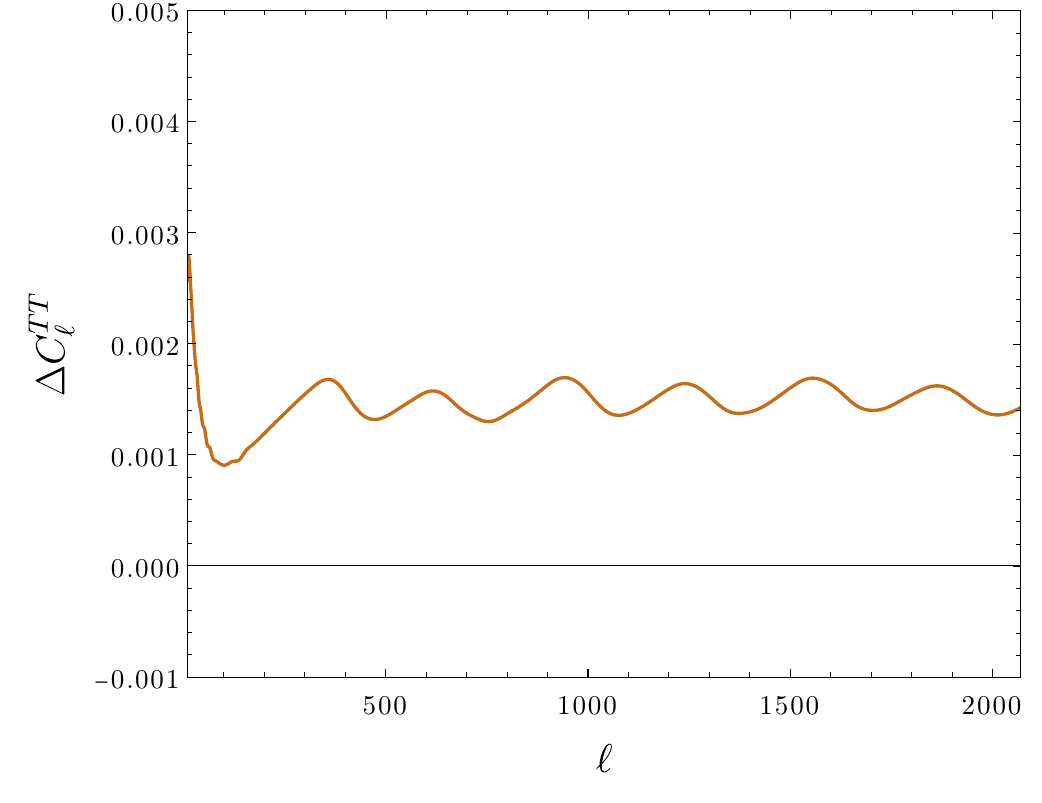}}
\subfigure[]{\includegraphics[height=5.4cm]{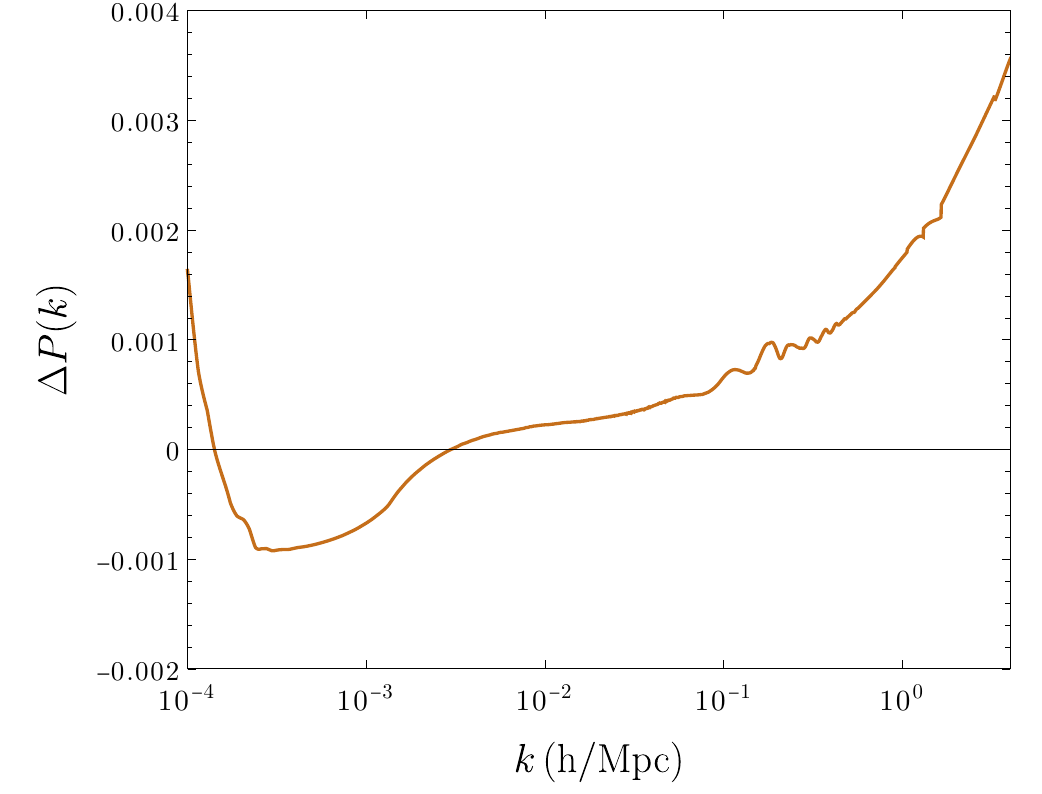}}
\caption{Same as in Fig.~\ref{fig:perturbations_benchmark}, but for the parameter values listed in Table~\ref{tab:68_confidence_intervals}. Parameters with only upper bounds are fixed at their maximal allowed values. Note the change in the vertical axis scale. The data have been omitted for clarity.}
\label{fig:perturbations}
\end{figure}

At the perturbation level, Fig.~\ref{fig:perturbations_benchmark} shows the induced changes in the CMB temperature and matter power spectra, $C_\ell$ and $P(k)$ respectively, for selected values of the ALP decay width $\Gamma_\phi=\{5,20,50\}~\textrm{Km/s/Mpc}$. The results are presented in terms of the fractional change $\Delta X$, where $X=C_\ell$ or $P(k)$, and the $\Lambda\rm CDM$ predictions are computed using the default parameters from the CLASS \textit{.ini} file, listed in Table~\ref{tab:base_lcdm_params}. The effects are qualitatively similar to those described in Ref.~\cite{McCarthy:2022gok}. The CMB temperature power spectrum exhibits an enhancement at low multipoles due to a stronger late-time integrated Sachs–Wolfe effect, arising from the time dependent gravitational potential experienced by CMB photons. This effect is mainly driven by the increased dark energy contribution required to maintain spatial flatness, together with the modified evolution of ALP DM and DR. At high multipoles, the spectrum is generally suppressed due to reduced CMB lensing resulting from the lower DM density. This weakens the smoothing of the acoustic peaks due to the lensing, making peaks more pronounced and troughs deeper relative to $\Lambda\rm CDM$. In addition, the reduced transfer of power from large to small scales leads to a mild overall suppression toward high-$\ell$.

In the linear matter power spectrum, we observe an enhancement with respect to $\Lambda\rm CDM$ at large scales, $k\lesssim10^{-2}~\rm h/Mpc$, followed by a suppression at smaller scales. The conversion of DM into DR suppresses structure growth and reduces CMB lensing, potentially leading to lower values of $S_8$ compared to $\Lambda\rm CDM$. On large scales, the modified expansion history shifts the matter-radiation equality scale, displacing the turnover in the matter power spectrum and enhancing power at low $k$.

On the other hand, using the results provided by Cobaya, for both ALP model and $\Lambda\rm CDM$ in Table~\ref{tab:68_confidence_intervals}, we obtain the plots in Fig.~\ref{fig:perturbations}. For the sake of clarity, we only show the deviation of our model with respect to $\Lambda\rm CDM$, omitting the data. Despite the reduction in structure growth and the associated decrease in $S_8$, the ALP decay width $\Gamma_\phi$ is tightly constrained by Planck CMB observations. The primary CMB spectrum remains sensitive to late-time cosmology through the integrated Sachs–Wolfe effect and, more importantly, through CMB lensing. Since the reduced DM abundance weakens structure formation, it also reduces the lensing-induced smoothing of the acoustic peaks, producing deviations in the observed high-$\ell$ CMB spectrum that are strongly constrained by Planck data. Nevertheless, for sufficiently small values of $\Gamma_\phi$, the deviations from $\Lambda\rm CDM$ remain at the sub-percent level, with relative changes in both $C_\ell$ and $P(k)$ of order $\mathcal{O}(0.5\%)$. Despite these small modifications, the model provides a slightly improved fit to the data, with $\Delta\chi^2 \equiv \chi^2_{\rm ALP}-\chi^2_{\Lambda\rm CDM}=-3.75$ for the full dataset and $\Delta\chi^2_{\rm CMB}=-0.63$ for CMB data alone.


\newpage

\section{Conclusions}\label{conclusions}

ALPs in cosmology are very versatile. In this article, we worked in the pre-inflationary scenario in which the global symmetry that gives rise to the ALP is broken before inflation, at the scale $f_\phi$. Inflation homogenizes the ALP field so that it behaves as a classical condensate during radiation domination epoch. Typically, the ALP potential takes a cosine form. At the beginning, the ALP is displaced with respect to one of these minima and is frozen due to the Hubble friction, thereby behaving as DE with $\omega_\phi=-1$. When the condition $m_\phi>3 H(a_{\rm osc})$ is fulfilled during radiation domination epoch, the ALP begins to oscillate around the minimum, and these oscillations can account for all the observed DM. This constitutes the so called standard misalignment scenario.

We extended the above scenario by introducing a pre-oscillatory phase with constant equation of state $\omega_\phi \in [-1,1]$, without modifying the standard cosmological history. This phase is described by a scalar potential, admitting tracking behavior when $\omega_\phi\leq 1/3$, while for $\omega_\phi>1/3$ it generically depends on the choice of initial conditions. It is followed by a transition in the ALP potential, where a sudden change in its slope connects to the usual cosine-like form, ultimately triggering oscillations during the radiation era. We remain agnostic about the origin of this transition. The transition is characterized by the parameter $r_\phi$, defined as the ratio of the ALP energy at the end of the phase with constant $\omega_\phi$ to the height of the cosine potential. The sudden change in the potential typically drives the ALP into a kinetic misalignment phase, whose duration is determined by $r_\phi$ and $a_{\rm osc}$, before the field becoming trapped around one of the minima of the cosine potential and starting to oscillate, provided that $m_\phi > 3H$.

On the other hand, cosmology is experiencing a high precision era. As a consequence, tensions have emerged between different datasets. The persistence of the $H_0$ and $S_8$ tensions, which can be alleviated by an unstable dark matter component, motivates the introduction of a coupling between the ALP and a dark radiation sector, $g_{\phi\rm DR}$, allowing for its decay. ALP oscillations source the emission of DR rendering the ALP unstable with a decay rate $\Gamma_\phi$. We have explicitly calculated the DR energy density using the WKB approach and the stationary phase approximation, treating the ALP as a background field and subsequently including the backreaction of the produced DR on the ALP through energy conservation.

Later, we focused on the perturbations. For the ALP, we calculated the effective sound speed of the fluid. During the pre-oscillatory phase, we recovered the standard $c_1=1$. To describe the rapid oscillations that occur after the kinetic misalignment phase induced by the change in the ALP potential, we averaged the pressure, energy density, and velocity divergence over one oscillation period, assuming a slowly varying cosmological evolution. We obtained an expression for $c_1$ in the oscillatory phase that reproduces the correct limits in both the large and small momentum regimes. In the case of DR, we derived the equations for the multipoles in the Boltzmann hierarchy.

We have implemented this model in the software CLASS and performed a global fit using Cobaya to explore its cosmological implications. The analysis combines the most recent likelihoods, including CMB temperature and polarization data from \textit{Planck}, CMB lensing, supernovae from DES and Pantheon+, BAO measurements from DESI, and a Gaussian prior on $H_0$ to account for the results from the SH0ES collaboration. As a result, we obtained robust constraints on the parameters of our model. In particular, the data show a preference for a negative ALP equation of state, $\omega_\phi<-0.037$, prior to the onset of the oscillations, and place an upper bound on the ratio of the ALP energy density before the transition and the height of the cosine potential, $\log_{10}r_\phi<11.1$. 
Our analysis also placed bounds on the scale $f_\phi$, namely, $f_\phi\in [80,1.5\times 10^{10}]$~TeV and on the value of the scale factor at the onset of the oscillations $\log_{10}a_{\rm osc}=-8.9^{+3.2}_{-2.6}$. Since the ALP accounts for all the DM, the parameters $f_\phi$, $a_{\rm osc}$, $m_\phi$ and the DM energy density, $\Omega_{\rm c}$, are related through Eq.~(\ref{relation mfa_osc}), which allowed us to obtain the band in the $m_\phi$--$f_\phi$ plane shown in the first panel of Fig.~\ref{fig:results}. Within the allowed range of $f_\phi$, the ALP masses lie in the range $m_\phi\in[10^{-20},10^{-2}]$~eV. It is remarkable that our model allows access to smaller values of $f_\phi$ for a given mass than in the standard misalignment scenario. 

Regarding the production of DR, our analysis places an upper bound on the ALP decay rate $\log_{10}\Gamma_\phi<-1.88$. This is not sufficient by itself to alleviate the $H_0$ and $S_8$ cosmological tensions. The preference for small decay rates can be understood in terms of their impact on perturbations, as shown in Fig.~\ref{fig:perturbations_benchmark}. In particular, the CMB spectrum is sensitive to late-time physics through CMB lensing, which is suppressed at high multipoles by the ALP decay and tightly constrained by Planck data. This constraint can be translated into bounds on the ALP coupling to dark radiation, $g_{\phi\rm DR}$, as a function of the ALP mass, as shown in the second panel of Fig.~\ref{fig:results}. Our bounds become more stringent than the perturbativity limit for ALP masses above $1~\mu\mathrm{eV}$.

Finally, with respect to the agreement with the data, the goodness of the fit is similar for both the ALP model and $\Lambda\rm CDM$. The ALP model yields a slightly better fit, with $\Delta\chi^2=\chi^2_{\rm ALP}-\chi^2_{\Lambda\rm CDM}=-3.75$ when all datasets in the global fit are included, and $\Delta\chi^2_{\rm CMB}=-0.63$ for the CMB alone. However, the improvement is not statistically significant, and both models remain in good agreement with the data. 

In our analysis, we have focused on a constant equation of state $\omega_\phi\in[-1,1]$ before the onset of the oscillations and on ALP couplings to new gauge bosons; however, future studies could extend this framework by considering time-dependent equations of state, non-canonical scalar fields with $\omega_\phi>1$ or $\omega_\phi<-1$, as well as couplings to new fermions.

\section*{Code availability}
The CLASS code used to produce the results of this work and the Markov chains analyzed are available from the author upon reasonable request.
\section*{Acknowledgments}
I thank Antonio Cuesta for useful discussions and comments during the elaboration of this manuscript.
This work made use of the computational resources of the cluster at the Instituto de Física Corpuscular (IFIC), Valencia, Spain.

\printbibliography

\end{document}